\documentclass[10 pt]{article}
\newcommand{\btex}{0}
\newcommand{\aastex}{0}
\newcommand{\preprint}{1}
\usepackage{amsmath,graphicx,natbib}
\if1\aastex
\shorttitle{The Fallible F-test}
\shortauthors{van Dyk, David A.}
\else
\fi

\begin{document}
\if1\btex\bibliographystyle{/home/vandyk/TeXfiles/BibTeX/natbib}\fi

\if0\aastex
    \newcommand\apj{ApJ}
    \newcommand\apjl{ApJ}
    \newcommand\aap{A\&A}
    \newcommand\nat{Nature}
\fi
\newcommand\dist{\buildrel\rm d\over\sim}
\newcommand\mis{^{\rm mis}}
\newcommand\com{^{\rm com}}
\renewcommand\c{^{\rm com}}
\newcommand\bin{^{\rm bin}}
\newcommand\obs{^{\rm obs}}
\newcommand\aug{^{\rm aug}}
\newcommand\Ya{\dot Y}
\newcommand\Yb{\ddot Y}
\newcommand\Yc{\dot{\ddot Y}\hskip -0.8pc\phantom{Y}}

\newcommand{\pd}{{\partial}}

\newcommand\lamc{\tilde{\cal F}}
\newcommand\alc{\alpha}
\newcommand\itoN{^N_{i=1}}
\newcommand\iton{^n_{i=1}}
\newcommand\jtoJ{^J_{j=1}}
\renewcommand\th{\theta}
\newcommand\ind{\stackrel{\rm indep.}{\sim}}
\newcommand\updot{^{\textstyle\cdot}}
\newcommand\downdot{_{\textstyle\cdot}}
\renewcommand\r{\right}
\renewcommand\l{\left}
\newcommand\Var{{\rm Var}}
\newcommand\E{{\rm E}}
\newcommand\cur{^{(t)}}

\newcommand{\btheta}{\boldsymbol{\theta}}
\newcommand{\bphi}{\mbox{\boldmath{$\phi$}}}
\newcommand{\bY}{\boldsymbol{Y}}
\newcommand{\by}{\boldsymbol{y}}
\newcommand\bx{\boldsymbol{x}}

\newcommand\spacingset[1]{\renewcommand{\baselinestretch}%
{#1}\small\normalsize}

\spacingset{1}

\title{\bf Statistics: Handle with Care\\
\large Detecting Multiple Model Components with the Likelihood Ratio Test}

\if0\aastex
\author{Rostislav Protassov and David A. van Dyk
  \thanks{
    The authors gratefully acknowledge
    funding for this project partially provided by NSF grants
    DMS-97-05157 and DMS-01-04129, and
    by NASA Contract NAS8-39073 (CXC).}\\
  Department of Statistics, Harvard University
\and
Alanna Connors\\
Eureka Scientific
\and
Vinay L. Kashyap and Aneta Siemiginowska \\
Harvard-Smithsonian Center for Astrophysics}

\maketitle
\fi

\if1\aastex

\author{Rostislav Protassov and David A. van Dyk}
\affil{Department of Statistics, Harvard University}
\affil{One Oxford Street, Cambridge, MA 02138}
\email{protasso@stat.harvard.edu vandyk@stat.harvard.edu}
\and
\author{Alanna Connors}
\affil{Eureka Scientific}
\affil{2452 Delmer Street Suite 100, Oakland CA 94602-3017}
\email{connors@frances.astro.wellesley.edu}
\and
\author{Vinay L. Kashyap and Aneta Siemiginowska}
\affil{Harvard-Smithsonian Center for Astrophysics}
\affil{60 Garden Street, Cambridge, MA 02138}
\email{kashyap@head-cfa.harvard.edu aneta@head-cfa.harvard.edu }
\fi

\begin{abstract}

The likelihood ratio test (LRT) and the related $F$ test,\footnote{The
  $F$ test for an additional term in a model, as defined in
  \cite{bevington:69} on pp. 208-209, is the ratio
  $$F_\chi=\frac{\chi^2(m)-\chi^2(m+1)}{\chi^2(m)/(N-m-1)}=
  \Delta\chi^2/\chi^2_\nu,$$
  \noindent
  where $\chi^2(m)$ and $\chi^2(m+1)$ are the values of $\chi^2$
  statistic resulting from fitting $m$ and $m+1$ free parameters
  respectively and $\chi^2_\nu$, in Bevington's (1969) notation,
  stands for a $\chi^2$ random variable with $\nu$ degrees of freedom
  divided by the number of degrees of freedom $\nu$. In the remainder
  of this paper we use $\chi^2_\nu$ to denote a $\chi^2$ random
  variable with $\nu$ degrees of freedom, as this notation is more
  standard.} popularized in astrophysics by \citet{eadie:etal},
\citet{bevington:69}, \citet{lampton:margon:bowyer:76},
\citet{cash:79}, and \citet{avni:etal} do not (even asymptotically)
adhere to their nominal $\chi^2$ and $F$ distributions in many
statistical tests common in astrophysics, thereby casting many
marginal line or source detections and non-detections into doubt.
Although the above references illustrate the many legitimate uses of
these statistics, in some important cases it can be impossible to
compute the correct false positive rate. For example, it has become
common practice to use the LRT or the $F$ test for detecting a line in
a spectral model or a source above background despite the lack of
certain required regularity conditions.  (These applications were not
originally suggested by \citet{cash:79} or by \citet{bevington:69}).
In these and other settings that involve testing a hypothesis that is
on the boundary of the parameter space, {\it contrary to common
  practice, the nominal $\chi^2$ distribution for the LRT or the $F$
  distribution for the $F$ test should not be used}.  In this paper,
we characterize an important class of problems where the LRT and the
$F$ test fail and illustrate this non-standard behavior.  We briefly
sketch several possible acceptable alternatives, focusing on Bayesian
posterior predictive probability-values.  We present this method in
some detail, as it is a simple, robust, and intuitive approach.  This
alternative method is illustrated using the gamma-ray burst of May 8,
1997 (GRB 970508) to investigate the presence of an Fe~K emission line
during the initial phase of the observation.

There are many legitimate uses of the LRT and the $F$ test in
astrophysics, and even when these tests are inappropriate, there
remain several statistical alternatives (e.g., judicious use of error
bars and Bayes factors).  Nevertheless, there are numerous cases of
the inappropriate use of the LRT and similar tests in the literature,
bringing substantive scientific results into question.
\end{abstract}

%\tableofcontents

\section{Introduction}

Distinguishing a faint spectral line or a new source from a chance
fluctuation in data, especially with low photon counts, is a
challenging statistical task.  As described in Section~2, these are
but two examples in a class of problems that can be characterized
in statistical terms as a
test for the presence of a component in a finite mixture distribution.
It is common practice to address such tests with a likelihood ratio
test (LRT) statistic or the related $F$ statistic and to appeal to the
nominal asymptotic distributions or {\it reference
  distribution}\footnote{As detailed below, the reference distribution
  is used to calibrate a test statistic. When choosing between two
  models, we assume the simpler or more parsimonious model holds and
  look for evidence that this assumption is faulty.  Such evidence is
  calibrated via the {\it reference distribution}, the known
  distribution of the test statistic under the simple model. If the
  observed test statistic (e.g., LRT or $F$ test) is extreme according
  to the reference distribution (e.g. $\chi^{2}_{1} > 10.83)$, the
  simple model is rejected in favor of the more complex model.} of
these statistics \cite[etc.]{mura:etal:88, feni:etal:88,   %%% Punctuation %%%
  yosh:etal:92, palm:etal:94, band:etal:95, band:etal:96,
  band:etal:97, freeman:etal:99, piro:etal:99}.  (See
\citet{band:etal:97} for a discussion of the close relationship
between the LRT and the $F$ test.)  The underlying assumption is that
in some asymptotic limit the statistic being used to describe the data
is distributed in an understandable way, and hence useful bounds may
be placed on the estimated parameters.  Unfortunately, the standard
asymptotic theory does not always apply to goodness-of-fit tests of
this nature even with a large sample size or high counts per bin.
Thus, use of these statistics may be misleading.  For example,
searches for cyclotron scattering and atomic lines in $\gamma$-ray
bursts based on such uncalibrated statistics may be unreliable.

In nested\footnote{E.g., the allowed parameter values of one model
must be a subset of those of the the other model.} significance tests
such as the LRT or $F$ test, we wish to choose between two models,
where one model (the null model) is a simple or more parsimonious
version of the other model (the alternative model).  We seek evidence
that the null model does not suffice to explain the data, by showing
the observed data are very unlikely under the assumption that the null
model is correct.  In particular, we define a test statistic (e.g.,
the LRT statistic, $F$ statistic, or $\Delta \chi^2$ values) with a
distribution that is known at least approximately assuming the null
model is correct.  We then compute the test statistic for our data and
compare the result to the known null distribution.  If the test
statistic is extreme (e.g., large) we conclude the null model is very
unlikely to produce such a data set and choose the alternative model.
A test statistic {\it without a known reference distribution} is
without standard justification and is of little direct use; we can
neither calibrate an observed value nor compute the false positive
rate.  Such is the case for the LRT statistic and the $F$ statistic
for detecting a spectral emission line, an absorption feature, or {\it
added} model component\footnote{We assume that when testing for the
presence of an emission line, model parameters are constrained so the
line intensity is greater than zero; similar assumptions are made for
absorption features and added model components.}.  Because this use is
outside the bounds of the standard mathematical theory, the reference
distribution is generally {\it uncalibrated: unknown and
unpredictable}.
%%%%%%% Trying to be very specific AND use shorter sentences. %%%%%%
This problem is fundamental, i.e., intrinsic to the definition of the
LRT and $F$ test.  It is not due to small sample size, low counts per
bin, or faint signal--to--noise ratio.  It persists even when
Gauss--Normal statistics hold and $\chi^2$ fitting is
appropriate.\footnote{As pointed out by the referee,
\citet{whea:etal:95} showed that least squares or $\chi^2$ fitting can
sometimes be equivalent to maximum likelihood fitting even when
Poisson statistics apply.  However, their method is not universally
applicable since it presumes that the weight matrix is independent of
(or only weakly dependent on) the Poisson means (see their Equations
12, 19, and 20), whereas in the Poisson case the weights are the
reciprocal of the Poisson means---if the weights are known, there is
nothing to estimate.}  Several authors \citep{mattox:etal:96,
deni:wald:99,freeman:etal:99} have recognized that the null
distribution of the LRT and $F$ statistics may vary from the nominal
tabulated values \citep[e.g., the tables given in][for the $F$
test]{bevington:69}.  Nonetheless, the inappropriate use of the $F$
test in the astrophysics literature is endemic.

%As an illustration, an electronic search for `F
%statistics', `F test', or `LRT' in papers published in ApJ, ApJL, and
%ApJS between 1995 and mid 2001 resulted in 183 hits.  These articles
%are categorized by how the test was used in Table~\ref{tbl:search}.
%This simple search resulted in over 127 articles that used the LRT or
%$F$ test in a {\it questionable} manner. We say questionable because
%many articles do not give sufficient details to determine if the test
%is used properly.  For example, we assume that when testing for the
%presence of an emission line, model parameters are constrained so that
%the fitted model will necessarily include an {\it added} emission
%line, as opposed to a subtracted, i.e., absorption line.  We believe
%this is common practice.  Likewise, we assume fitted absorption
%features are constrained to (locally) reduce the expected flux.
%Although for strong features such constraints are unlikely to effect
%the fitted model, they may be important for weak features where
%statistical tests are most useful in feature detection.  In any case,
%constraints have an important effect on the null distribution, which
%is computed under the assumption of no model feature.

%Table~\ref{tbl:search} gives an indication as to when the LRT and $F$
%test are or are not appropriate.  

As a rule of thumb, there are two important conditions that must be
satisfied for the proper use of the LRT and $F$ statistics.  First,
{\it the two models that are being compared must be nested}. Second,
{\it the null values of the additional parameters may not be on the
boundary of the set of possible parameter values}.  The second
condition is violated when testing for an emission line because the
line flux must be non-negative and the null value of this parameter is
zero, which is the boundary of the non-negative numbers.

Because of the first condition, it is, for example, inappropriate to
use the $F$ test to compare a power law with a black body model or to
compare a power law model with Raymond Smith thermal plasma model.
This issue is discussed in \citet{freeman:etal:99} and is not the
primary focus of this paper.  Instead, we focus on the second
condition that disallows, for example, testing for an emission line,
an absorption feature, or other {\it added} spectral components (e.g.,
a power law, Compton reflection component, black body component,
thermal component, etc.) or testing for a quasi-periodic oscillation
in timing data, an added Gaussian in light curves, or an {\it added}
image feature (e.g., an added point source).  We emphasize that there
are many legitimate uses of the LRT and $F$ test, e.g., testing for a
broken power law; comparing the variances of two samples; determining
if a spectrum is the same or different in two parts of an image, at
two difference times, or in two observations; or deciding whether to
allow non-solar abundances. Generally, returning to the two
rule-of-thumb conditions should guide one as to when the $F$ test and
LRT are appropriate.  (We note, however, that the reference
distributions of these tests are only reliable with a sufficiently
large data set, even when the two conditions are met.)

In the remainder of the paper we explain why these standard tests fail
and offer alternatives that work.  In Section~2, we look at the class
of models known as finite mixture models (which allow for multiple
model components).  We discuss the yet unresolved question of
determining the number of components in the mixture and show that
testing for a spectral line, a new source, or other {\it added} model
component are special cases of this problem.  The LRT and the $F$ test
have often been proposed as simple solutions in these cases. The LRT
is specifically discussed in Section~3, and is shown to be invalid
(i.e., uncalibrated) {\it in this setting} since as we discussed above
its basic criteria are not met. In Section~4, we discuss a number of
possible alternatives to the LRT and $F$ statistics including Bayes
factors, the Bayesian Information Criterion, and posterior predictive
p-values.\footnote{A probability-value or p-value is the probability
of observing a value of the test statistic (such as $\chi^2$) as
extreme or more extreme than the value actually observed given that
the null model holds (e.g. $\chi^2_{30} \geq 2.0$) Small p-values are
taken as evidence against the null model; i.e., p-values are used to
calibrate tests.  Posterior predictive p-values are a Bayesian
analogue; see Section~4.2.}  Complete recipes for all alternatives to
the LRT and $F$ test are beyond the scope of this paper.  As discussed
in Section~4, we focus on posterior predictive p-values because they
are conceptually and computationally simple and closely mimic the
nested significance tests described above; see the high--redshift
quasar example in Sections~3 and 4.  With this machinery in place, we
investigate a typical example in Section~5; we investigate whether the
data support the presence of an Fe K emission line during the initial
phase of the GRB 970508.  Here neither the LRT nor the $F$ test are
appropriate.  On the basis of our analysis, the model with a spectral
line is clearly preferable. In Section~6, we conclude with several
cautions regarding the inappropriate use of statistical methods which
are exemplified by the misuse of the LRT.

Throughout the paper we use the LRT for a spectral line as an example,
but the discussion and results apply equally to related tests (e.g.,
the $F$ test) and to other finite mixture models (e.g. how many
sources are detected above background; \citet{avni:etal}) or even more
generally to testing any null models on the boundary of the parameter
space.

\section{Finite Mixture Models in Astrophysics}

We begin with an illustration using a simple spectral model.  Consider
a source spectrum that is a mixture of two components: a continuum
modeled by a power-law $\alpha E^{-\beta}$ and an emission line
modeled as a Gaussian profile with a total flux $\tilde{\cal F}$.
% Start here Modified by AS/ Jan 26
%Formally, the expected counts,
%${\cal F}_j$, in bin $j$ with energy $E_j$ for a ``perfect'' instrument
%with effective area everywhere equal to the maximum possible effective
%area and instrument response equal to a delta function\footnote{We use
%  the maximum value of the effective area over the spectral energy
%  range of interest and ignore instrument response in this stage of
%  the analysis. This is only a matter of convenience and the full
%  effective area and instrument response variations are included in
The expected observed flux ${\cal F}_j$ from the source within an
energy bin $E_j$ for a ``perfect'' instrument is given by
%Equation~\ref{eq:defxi}.}
\begin{equation}
{\cal F}_j = dE_j \alpha E_j^{-\beta} + \tilde{\cal F} p_j
 \qquad {\rm for} \quad j=1,\ldots,J,
\label{eq:deflam}
\end{equation}
where $dE_j$ is the energy width of bin $j$, $\alpha$ is the
normalization constant for the power law with an index $\beta$, and
$p_j$ is the proportion of the Gaussian line that falls in bin $j$.
For simplicity we parameterize the Gaussian line in terms of its
location, $\mu$, and assume that its width, $\sigma$, is fixed.

To calculate the number of counts observed with a counting detector
that is
predicted by this model we need to take into account several
instrumental effects.  The observed spectrum is usually contaminated
with background counts, degraded by instrument response, and altered
by the effective area of the instrument and interstellar absorption.
Thus, we model the observed counts in a detector channel $l$
as independent Poisson\footnote{Recall, a random variable $X$ is said
  to follow a Poisson distribution with parameter or intensity ${\cal
    F}$ if $\Pr(X=x)=e^{-{\cal F}}{\cal F}^x /x!$.  In this case
  $\E(X)={\cal F}$ and we often write $X\dist{\rm Poisson}({\cal F})$
  (read as $X$ is distributed as Poisson with intensity ${\cal F}$).
  This representation conditions on the intensity parameter, ${\cal
    F}$, which in turn may vary.}  random variables with expectation
\begin{equation}
\xi_l = \sum_{j=1}^J R_{lj} A_j {\cal F}_j e^{-\gamma/E_j} + b_l,
\quad l=1,\ldots,L,
\label{eq:defxi}
\end{equation}
where $(R_{lj})$ is the photon redistribution matrix of size $L \times
J$ that summarizes the instrument energy response, $A_j$ is the
effective area at energy $E_j$ (normalized for convenience so that
$\max_j A_j = 1$), $\gamma$ is the parameter for simple exponential
absorption,\footnote{For mathematical transparency, we use a simplified
  exponential absorption throughout the paper except in the analysis
  of GRB 970508 where standard interstellar absorption is used.}
and $b_l$ is the expected background counts in channel $l$.  We
focus on the task of determining whether the data support the presence
of a Gaussian line as in Equation~\ref{eq:deflam}.

The form of the model specified by Equation~\ref{eq:deflam} is a
special case of what is known in the statistics literature as a finite
mixture model.  The simplest example of a finite mixture model is a
population made up of $K$ subpopulations which differ in their
distribution of some quantity of interest, $x$.  The distribution of
$x$ in subpopulation $k$ may have unknown parameters and is
represented by $p_k(x)$. The relative size of subpopulation $k$ is
$\omega_k \geq 0$ with $\sum_k \omega_k = 1$.  If we sample randomly from the
larger population, the distribution of $x$ is
\begin{equation}
f(x) = \sum_{k=1}^K \omega_k p_k (x).
\label{eq:defmix}
\end{equation}
Finite mixture models are an important
class of statistical models with applications in the social,
biological, engineering, and physical sciences.  A general review of
the topic can be found in any of the several useful books describing
these models, their applications, and their statistical properties
\citep*{ever:hand:81, titt:smit:mako:85, mcla:badf:88, lind:95}.

As an example in high energy spectral analysis, consider again
Equation~\ref{eq:deflam}.  We postulate that there are two ``types''
or ``subpopulations'' of photons, those originating from the continuum
and those originating from the Gaussian line.  The former have
distribution in the shape of a power law, the latter as a Gaussian
distribution.  The relative sizes of the two populations are
determined by $\alpha$ and $\tilde{\cal F}$.

There are many other statistical problems in astrophysics that can be
phrased in terms of determining $K$ (i.e., the true number of
components or subpopulations) in a finite mixture.  For example, consider
testing for the presence of a point source in spatial data or a burst
in time series data.  In both cases, $p_1(x)$ represents the
background or steady state model (i.e., model without a point source
or without a burst), and we wish to determine if the data support a
second component, $p_2(x)$, which represents the point
source or burst.  Other examples include testing for the presence of a
second plasma component in coronal temperature models of late-type
stars \citep{schm:etal:90, kaas:etal:96, sing:etal:99}.  In this case,
$p_k(x)$ in Equation~\ref{eq:defmix} represents a plasma
component for $k=1,2$.  Here the key is to determine if a single
plasma component (i.e, $K=1$) suffices to explain the data.

Although the simplified form given in Equation~\ref{eq:deflam} is used
for illustration throughout the paper, the difficulty with the LRT (or
the $F$ test) that is described in the next section applies equally to
the above problems, as does the Bayesian solution suggested in
Section~4.  It is well known in the statistical literature that
determining the number of components, $K$ (e.g., whether to include
the Gaussian line in Equation~\ref{eq:deflam}) is a challenging
problem that defies a standard analytical solution.  Notice that a
formal test of whether we should include component $k$ corresponds to
testing whether $\omega_k=0$, which is the {\it boundary} of the set
of possible values of $\omega_k$.  This issue is discussed at length
in the several texts referenced above.

\section{The Fallible Likelihood Ratio Test}

\subsection{Mathematical Background}

As discussed in Section~1, the LRT and the related $F$ test have often
been used to test for the presence of a spectral line.  To understand
the difficulty with using these tests in this setting, we begin with a
formal statement of the asymptotic result that underlies the LRT.

Suppose $\bx=(x_1,\ldots,x_n)$ is an independent sample (e.g.,
measured counts per PHA channel or counts per imaging pixel) from the
probability distribution $f(x|{\btheta})$, with parameters
${\btheta}=(\th_1\ldots,\th_p)$. We denote
the likelihood ratio statistic, $T_{\rm LRT}(x) = -2\log R(\bx)$,
where
\begin{equation}
R(\bx)={{\max \prod\iton f(x_i|\th_1^T,\ldots,\th_q^T,\th_{q+1},\ldots,\th_p)} \over
{\max  \prod\iton f(x_i|\th_1,\ldots,\th_p)}},
\label{eq:defLRTa}
\end{equation}
where the maxima are found by varying the parameters. In the
numerator, the $\th^T$ terms represent parameters which are not varied
but held at their `true' values, i.e., the values assumed under the
null model.  Under the {\it regularity conditions}\footnote{The
  details of these regularity conditions are quite technical in nature
  and are summarized in Appendix A.
  A mathematically rigorous statement can be found in
  \citet[Sections 4.2 and 4.4,][]{serf:80}.} discussed below, if
$(\th_1,\ldots,\th_q)$ actually equals $(\th_1^T,\ldots,\th_q^T)$, the
distribution of the likelihood ratio statistic converges to a $\chi^2$
distribution with $q$ degrees of freedom as the sample size,
$n\rightarrow\infty$. Equation~\ref{eq:defLRTa} can be written more
formally by defining $\Theta$ to be the set of possible parameters
values of ${\btheta}$. We are interested in testing whether ${\btheta}$ is
in some subset of the parameter set, $\Theta_0 \subset \Theta$.  We
then denote the likelihood ratio statistic, $T_{\rm LRT}(x) = -2\log
R(\bx)$, where
\begin{equation}
R(\bx)={{\sup_{{\btheta} \in \Theta_0} L({\btheta} |\bx)} \over
{\sup_{{\btheta} \in \Theta} L({\btheta} |\bx)}}
\label{eq:defLRT}
\end{equation}
with $L({\btheta} | \bx)$ denoting the likelihood function,
$\prod_{i=1}^n f(x_i|{\btheta})$.  Again, under the regularity
conditions, if ${\btheta} \in \Theta_0$, the distribution of the
likelihood ratio statistic converges to a $\chi^2$ distribution as the
sample size, $n\rightarrow\infty$.  The degrees of freedom of the
$\chi^2$ distribution is the difference between the number of free
parameters specified by $\Theta$ and by $\Theta_0$. (In
Equation~\ref{eq:defLRT} we replace the `max,' i.e., maximum, in
Equation~\ref{eq:defLRTa} with `sup,' i.e., supremum. This is a
technicality which is mathematically necessary when the maximum occurs
on the boundary of the parameters space. E.g., $\max_{x>0} 1/(x+1)$ is
not defined, but $\sup_{x>0} 1/(x+1) = 1$.)

Examples of the LRT are found widely in scientific applications; see
e.g., the many citations in Section~1.  As a simple example, suppose
one wishes to test whether the spectrum of a source is the same in two
spatially separated regions.  For simplicity, suppose a simple power
law with two parameters (normalization and power law index) is used to
model the spectrum.  In the denominator of Equation~\ref{eq:defLRT},
we fit the power law independently in the two regions via maximum
likelihood and multiply the two likelihoods evaluated at there
respective maximum likelihood estimates.  For the numerator, we simply
fit the combined data set via maximum likelihood.  The resulting test
statistic, $-2\log R(\bx)$ is approximately $\chi^2_2$, i.e., $\chi^2$
in distribution with two degrees of freedom, the difference in the
number of free parameters in the two models.
%%%See \chi^2 notation note! %%%

As noted by \citet{cash:79}, this is a remarkably general result with
very broad application.  It must, however, be used cautiously; it is
not a universal solution and does not apply in all settings.
(This was also noted by \citet{freeman:etal:99}.)  In
particular, the ``regularity conditions'' mentioned above are
concerned mainly with the existence and behavior of derivatives of
$\log L({\btheta}|\bx)$ with respect to ${\btheta}$ (e.g., $\log
L({\btheta}|\bx)$ must be three times differentiable), the topology of
$\Theta$ and $\Theta_0$ (i.e., $\Theta_0$ must be in the interior of
$\Theta$ and both must be open), and the support of the distribution
$f(\bx|{\btheta})$ (i.e., the range of possible values of $x$ cannot
depend on the value of ${\btheta}$).  Details of these regularity
conditions are given in Appendix A. The difficulty with testing for a
spectral line or more generally for the existence of a component in a
finite mixture model lies in the topology of $\Theta$. The standard
theory requires that $\Theta$ be an open set---the LRT cannot be used
to test whether the parameter lies on the boundary of the parameter
space.  Consider the source model given in Equation~\ref{eq:defxi}. If
we set ${\btheta}=(\alpha, \beta, \gamma, \tilde{\cal F})$ and test
$\tilde{\cal F}=0$ we are examining the boundary of the parameter
space, and the LRT statistic is not (even asymptotically) $\chi^2$ in
distribution.

In Appendix~B, we show mathematically that the distribution of the LRT
is not $\chi^2$ when testing for a spectral line. This is further
illustrated via a simulation study in the next section.

\subsection{Computing the False Positive Rate for Fully Specified Models}

Unfortunately, analytical results for the sampling distribution of the
LRT do not exist in many settings, such as testing for a spectral
line.  Thus, we resort to two simulation studies which illustrate the
unpredictable behavior of the distribution of the LRT statistic.

\paragraph{Simulation 1: Testing For an Emission Line.}

The most common problem in spectral analysis of quasar X-ray data is
detection of a fluorescent emission line, Fe-K$\alpha$ at $\sim
6.4-6.8$~keV.  The iron line indicates existence of a cold material in
the nucleus and its energy and width can constrain the ionization
state of the matter and its distance from the central black hole. The
line is strong and easily detected in some low redshift, low luminosity
sources. However, for high luminosity and high redshift objects its
presence in the data is not obvious and the common statistical
technique used to support the detection of this line is the $F$ test.

Our first simulation is designed to mimic the ASCA/SIS observation of
the high redshift $(z=3.384)$ \if1\aastex\objectname{quasar S5
  0014+81} \else quasar S5 0014+81\fi \citep{elvi:etal:94}.  We use
standard ASCA response matrices in the simulations.\footnote{
ftp://legacy.gsfc.nasa.gov/caldb/data/asca/sis/cpf/94nov9}  The
effective area file was created with FTOOLS (ascaarf v.2.67) and the
corresponding background file was extracted from the ASCA Background
Blank Fields.\footnote{
ftp://legacy.gsfc.nasa.gov/caldb/data/asca/sis/bcf/bgd/94nov}

We assume that the quasar emission in the observed energy range is
described by the model given in Equation~\ref{eq:deflam} with a power
law continuum $(\alpha=1.93\times 10^{-3}$ counts per cm$^2$ per keV per
second, $\beta=2.11)$, and exponential absorption parameter $(\gamma=
1.69)$. These parameter values correspond to fitted values for the
quasar observation obtained by \citet{vand:00}.

We simulated 200 data sets,
$\{\tilde{\bx}^{(t)},t=1,\ldots,200\}$
%using background, effective
%area, and instrument response corresponding to the quasar observation
%{\bf how should we say this? refer to ASCA??},
%assuming a power law continuum $(\alpha=1.93e-3,\beta=2.11)$, and
%exponential absorption parameter $(\gamma= 1.69)$.
%The parameter
%values correspond to fitted values for the quasar observation.
and each was fit three times via maximum likelihood
using the EM algorithm as described by \citet{vand:00} using three
different models:
\begin{description}
\item[Model 1:] continuum (i.e., power law plus exponential
  absorption);
\item[Model 2:] continuum plus a narrow line with one free parameter,
  $\tilde{\cal F}$, and location fixed at 1.55~keV, and width
%standard deviation
fixed at zero;
\item[Model 3:] continuum plus a wide line with two free parameters,
  $\tilde{\cal F}$ and location, with the width fixed at $\sigma=0.28~{\rm keV}$.
\end{description}
%The line in the final model is much wider than we would expect to fit
%in practice, but this allows it to move freely across the continuum
%and thus suits our simulation well.

For each data set, we computed two LRT statistics,
$T_2(\tilde{\bx}^{(t)})=-2\log(L(\hat{{\btheta}}_1|\tilde{\bx}^{(t)})
/L(\hat{{\btheta}}_2|\tilde{\bx}^{(t)}))$ and $T_3(\tilde{\bx}^{(t)})=
-2\log(L(\hat{{\btheta}}_1|\tilde{\bx}^{(t)})/L(\hat{{\btheta}}_3|
\tilde{\bx}^{(t)}))$, where $\hat{{\btheta}}_1, \hat{{\btheta}}_2$, and
$\hat{{\btheta}}_3$ are the maximum likelihood estimate under Models 1,
2, and 3, respectively.  Since the data were simulated under the null
model (i.e., Model 1), the histograms of the computed LRT
statistics in the first two panels of Figure~\ref{fig:lrt} represent
the reference distribution of the LRT against these two alternative
models.  The nominal $\chi^2$ distributions with one and two degrees
of freedom are plotted on the histograms and clearly do not suffice.
The false positive rates are 2.6\% and 1.5\% in the nominal 5\% tests,
respectively.  In this case, the LRT understates the evidence for an
emission line. Correcting the false positive rate should enable us to
detect weak lines that would be missed by blind application of the
LRT.

\if1\aastex\placefigure{fig:lrt}\fi

\paragraph{Simulation 2: Testing For a Simple Absorption Line.}
Although the LRT is conservative in both of the tests in Simulation~1,
this is not always the case. This can be seen in a second simulation
where we consider a simplified absorption line.  Although
multiplicative model components such as an absorption line do not
correspond to testing for a component in a finite mixture, the LRT
still does not apply if the null model is on the boundary of the
parameter space; such is the case with absorption lines.  In this
simulation we ignore background contamination, instrument response,
and binning.  We simulate 1000 data sets each with 100 photons from an
exponential continuum, and fit two models:
\begin{description}
\item[Model 1:] exponential continuum;
\item[Model 2:] exponential continuum plus a two parameter
                absorption line, where the fitted absorption probability is
                constant across the line which has fixed width,
                but fitted center.
\end{description}
Again, we computed the LRT statistic for each of the 1000 simulated
data set and plotted the results in the final panel of
Figure~\ref{fig:lrt}.  Clearly the LRT does not follow its nominal
reference distribution ($\chi^2$ with 2 degrees of freedom) even with
this simplified absorption line model; the false positive rate is
31.5\% for the nominal 5\% test.  That is, use of the nominal
reference distribution would result in over six times more false line
detections than expected.

\section{Bayesian Model Checking}

Although some theoretical progress on the asymptotic distribution of
$T_{\rm LRT}(\bx)$ when $\Theta_0$ is on the boundary of $\Theta$ has
been made (e.g., by \citet{cher:54} and specifically for finite
mixtures by \citet{lind:95}), extending such results to a realistic
highly structured spectral model would require sophisticated
mathematical analysis.  (See \citet{lind:95} for a simple exception
when only $\omega$ is fit in Equation~\ref{eq:defmix}.)  In this
section, we pursue a mathematically simpler method based on Bayesian
model checking known as posterior predictive p-values
\citep*{meng:94a,gelm:meng:ster:96}.  As we shall see, this Bayesian
solution is simpler and far more generally applicable than the
asymptotic arguments required for satisfactory behavior of the LRT.

Posterior predictive p-values are but one of many methods for model
checking and selection that may be useful in astrophysics. Our aim
here is not to provide a complete catalog of such methods, but
rather to provide practical details of one method that we believe is
especially promising and little known in astrophysics. In
Section~4.3, we provide a brief comparison with several other Bayesian
methods.

\subsection{The Posterior Predictive P-value}
\label{subsec:ppp}
The central difficulty with the LRT and the $F$ test in this setting
is that their reference distributions are unknown even asymptotically.
Moreover, the distributions likely depend on such things as the
particular shape of the continuum, the number of lines, and their
profiles and strengths.  Thus, it is difficult to obtain any general
results regarding such reference distributions even via simulation.
The method of posterior predictive p-values uses information about the
spectrum being analyzed to calibrate the LRT statistic (or any
other test statistic) for each particular measurement.  In the
simulations described in Section~3.2, we simulated data sets,
$\tilde{\bx}^{(t)}$, using a fixed value of $(\alpha,\beta,\gamma)$
and observed the behavior of $T_{\rm LRT}(\tilde{\bx}^{(t)})$.
Instead of fixing the model parameter at its fitted value under the
null model, the method of posterior predictive p-values uses values of
the parameter that are relatively likely given the observed counts.
That is, {\it we run a Monte Carlo simulation to access the sampling
distribution of the LRT (or other) statistic so that we can calibrate
the value of the statistic computed on the data and determine a
p-value.}  The Monte Carlo simulation is run using parameter values
fit to the data under the null model and {\it accounts for uncertainty
(i.e., error bars)} in these fitted values.
%%%% ? Look this up ?
%{\bf Ref to Eadie et al, assuming Gauss-Normal error bars??
%Vinay have you the reference?}

%%% Note on usage %
To formalize this, we review
Bayesian model fitting which we use to simulate
values of the parameter via Monte Carlo.
%%Slightly more specificity:%%
Bayesian model fitting involves the probability of the parameters,
given the model: $p({\btheta} | \bx, I)$; while ``classical'' or
``frequentist'' statistics deals with the converse: $p(\bx|{\btheta},
I)$.  Interested readers are referred to the recent more detailed
treatments of Bayesian methods given in
\citet*{gelm:carl:ster:rubi:95}, \citet*{carl:loui:96}, and
specifically for astrophysics, in \citet*{vand:etal:01}.
%%% Again a little more specificity: $$$$
One transforms from $p(\bx|{\btheta}, I)$ to $p({\btheta} | \bx, I)$
or visa versa using Bayes theorem, which states
\begin{equation}
p({\btheta} | \bx, I) = {p(\bx|{\btheta}, I) p({\btheta} | I) \over p(\bx|I)},
\label{eq:bayesthm}
\end{equation}
where $p({\btheta}|\bx, I)$ is the distribution of the parameter given
the data, and any available prior information, $I$, i.e., {\it the
posterior distribution}, $p(\bx|{\btheta}, I)$, is the model for the
distribution of the data $\bx$, i.e., {\it the likelihood},
$p({\btheta}| I)$ contains information about the parameter known prior
to observing $\bx$, i.e., the {\it prior distribution}, and $p(\bx|I)$
is a normalizing constant for $p({\btheta}|\bx,I)$.
Equation~\ref{eq:bayesthm} allows us to combine information from
previous studies or expert knowledge through the prior distribution
with information contained in the data via the likelihood.  In the
absence of prior information we use a relatively non-informative prior
distribution.  \citet{vand:etal:01} describe how to use MCMC and the
Gibbs sampler to simulate from $p({\btheta}|\bx)$ using spectral
models which generalize Equation~\ref{eq:deflam}.

Our procedure is given for an {\it arbitrary} statistic $T(\bx)$
(i.e., an arbitrary function $T(\bx)$ of data $\bx$) but certainly is
valid for $T_{\rm LRT}(\bx)$, i.e., the LRT statistic or the
$F$ statistic.  We calibrate
$T(\bx)$ using the distribution of $T(\tilde{\bx}^{(t)})$ given the
observed data $\bx$, as the reference distribution, i.e.,
\begin{equation}
p(T(\tilde{\bx}^{(t)})|\bx) = \int
p(T(\tilde{\bx}^{(t)}),{\btheta}|\bx) d{\btheta} = \int
p(T(\tilde{\bx}^{(t)})|{\btheta})p({\btheta}|\bx)d{\btheta},
\label{eq:distT}
\end{equation}
where the second equality follow because $\bx$ and
$\tilde{\bx}^{(t)}$ are independent given ${\btheta}$. (Here and in
what follows we suppress explicit conditioning on the prior
information, $I$.) What is important in Equation~\ref{eq:distT} is that
{\it we do not fix $\btheta$ at some estimated value in the reference
  distribution}; rather we integrate over its uncertainty as
calibrated by the posterior distribution.

Although analytical results are typically not available, calibration
is easily accomplished via Monte Carlo.  Specifically, we

\begin{enumerate}
\item Simulate parameter values $\{{\btheta}^{(t)},t=1,\ldots, N\}$ from
  $p({\btheta}|\bx)$, e.g., using the method of \citet{vand:etal:01}.
\item For $t=1,\ldots,N$ simulate $\tilde{\bx}^{(t)} \dist
  p(\bx|{\btheta}^{(t)})$, i.e., according to the model simulate $N$
  data sets, one for each simulation of the parameters obtained in Step 1.
  (This is similar to the often used ``parametric bootstrapping'', but
  now uncertainties in the parameters are accounted for.)
\item For $t=1,\ldots,N$ compute the statistic $T(\tilde{\bx}^{(t)})$.
  For $T_{\rm LRT}(\tilde{\bx}^{(t)})$ this involves computing the
  maximum likelihood estimates for the null and alternative models
  using each of the $N$ data sets, see Equation \ref{eq:defLRT}.
\item Compute the posterior predictive p-value,
\begin{equation}
p={1 \over N} \sum_{t=1}^N {\cal I}\{T(\tilde{\bx}^{(t)}) > T (\bx)\},
\end{equation}
where ${\cal I}\{\rm statement\}$ is an indicator function which is
equal to one if the statement is true and zero otherwise.
\end{enumerate}

The posterior predictive p-value is the proportion of the Monte Carlo
simulations that result in a value of $T(\tilde{\bx}^{(t)})$ more extreme
than the value computed with the observed data, $T(\bx)$.  If this is
a very small number, we conclude that our data is unlikely to have
been generated under the posterior predictive distribution.
Since this distribution is
computed assuming the null model,
we reject the null model and investigate alternative models.
That is, $p$ is treated as a p-value with small values indicating
evidence for the more complex model, e.g., the model with an
additional spectral line.

\subsection{An Example}
Here we illustrate the method of posterior predictive p-values by
testing for a spectral line in data obtained from a high redshift
quasar used in Simulation~1 of Section~3.2.  In particular, we compare
Model~3 with Model~1 as described in Section~3.2.  Calculations
proceed exactly as in Section~3.2, except each $\tilde{\bx}^{(t)}$ is
simulated with a different ${\btheta}^{(t)}$ as simulated from
$p({\btheta}|\bx)$.
%% Unusual usage:
Five hundred simulations from the resulting posterior predictive
distribution of $T_{\rm LRT}(\bx)$ are displayed in
Figure~\ref{fig:ppp} along with a vertical line that correspond to
$T_{\rm LRT}(\bx)$ computed with the observed data.  The resulting
posterior predictive p-value is 1.6\% showing strong evidence for the
presence of the spectral line.

\if1\aastex\placefigure{fig:ppp}\fi

\subsection{General Advice on Model Checking and Model Selection}

Posterior predictive p-values are by no means the only statistical
method available for model checking and model selection; in this
section we outline some important Bayesian alternatives and discuss
our recommendation of using posterior predictive p-values in this
case. This material may seem somewhat technical to some readers, but
may be skipped since the remainder of the paper is independent.

We begin by emphasizing that detecting model features is a challenging
statistical problem indeed; there is no consensus within the
statistical community as how best to proceed. Whenever possible, the
issue should be avoided by for example focusing on estimating the
perhaps very weak strength of a spectral line rather that deciding
whether there is a line. Practically speaking, however, we must decide
which and how many lines to include in the final analysis and would
like statistical tools that offer guidance.

Posterior predictive p-values aim to point out when the null model is
inadequate to describe the observed data; more counts than expected
under the null model in a narrow energy range is evidence that the
null model is missing a spectral line.  Since posterior predictive
p-values approximate the reference distribution of the test statistic
by simulating data using the null model, they tend to favor the null
model. That is, posterior predictive p-values tend to be conservative,
especially if the test statistic is poorly suited for detecting the
model feature in question \citep{meng:94a, baya:berg:99}. Nonetheless,
as illustrated in Section~5 posterior predictive p-values can be used
to detect model features and their conservative nature adheres to the
scientific standard for burden of proof.

To elicit more power from the posterior predictive p-value,
\citet{baya:berg:99} suggest concentrating on the parameters of
interest.  (More specifically, this method conditions on sufficient
statistics for the nuisance parameters when computing the p-value.)
Although less conservative than posterior predictive p-values, this
method is mathematically and computationally more demanding.

Bayes factors are a popular method for Bayesian model selection; here
we briefly compare them with posterior predictive analysis to explain
our preference for the latter, at least when well specified prior
information is not forthcoming.  The primary difficulty with Bayes
factors is that relative to posterior predictive p-values,
they are much more sensitive to the prior distribution.
Consider again testing
for a spectral line by selecting between two models
\begin{description}
\item[{\bf Model 1:}] a power law with no emission line;
\item[{\bf Model 2:}] a power law with an emission line.
\end{description}
Suppose that Model~1 has two free parameters and Model~2 has four free
parameters, the line intensity and location in addition to the power law
parameter and normalization. The Bayes factor is defined as
\begin{equation}
B={\int p(\bY|{\btheta}_1,{\rm Model \ 1})p({\btheta}_1|{\rm Model \ 1})d{\btheta}_1
\over \int p(\bY|{\btheta}_2,{\rm Model \ 2})p({\btheta}_2|{\rm Model \ 2})d{\btheta}_2},
\label{eq:bayfac}
\end{equation}
where $\bY$ are the counts, ${\btheta}_1$ are the parameters for Model~1
and ${\btheta}_2$ are the parameters for Model~2. The Bayes factor is
equivalent to the ratio of the posterior and prior odds for Model~1
versus Model~2.  Roughly speaking, large values of $B$ indicate that
the data favor Model 1 and small values that the data favor Model 2.
Computing a Bayes factors involves marginalizing over
(or averaging over) all unknown parameters
and thus can be computationally demanding; methods appear
elsewhere \citep{connors:97, freeman:etal:99, greg:lore:92, kass:raft:95}.

In a paper describing the natural ``Ockham's Razor" property of Bayes
factors, \cite{berg:jeff} point out that a Bayes factor goes (roughly)
like the height of the likelihood at its maximum times its width,
divided by the width of the prior.  Hence Bayes factors are strongly
dependent on the prior distributions for the parameters. This can be
seen formally in Equation~\ref{eq:bayfac}, where the numerator is the
{\it prior predictive distribution} under Model 1; likewise for the
denominator.  The prior predictive distribution also appears as the
denominator in Bayes theorem as a normalizing constant. This
distribution quantifies the distribution of the data with uncertainty
in the model parameters as described by the prior distribution. Thus,
if a highly diffuse prior distribution is used the prior predictive
distribution will also be very diffuse.  If the prior distribution is
improper,\footnote{An improper distribution is a distribution that is
{\it not} integrable and thus is not technically a distribution. One
should use improper prior distributions only with great care since in
some cases they lead to improper posterior distributions which are
uninterpretable.} neither the prior predictive distribution nor the
Bayes factor are defined. (There have been several attempts to define
analogous quantities in this situation; see \cite{smi:spie:80} and
\cite{kass:raft:95}.)  As a result, Bayes factors are {\it very}
sensitive to the choice of prior distribution; the more diffuse the
prior distribution for the line parameters the more the Bayes factor
will favor Model 1. Thus, when using Bayes factors the prior
distributions must be proper and well specified. That is, the prior
distribution must not be chosen in an ad hoc fashion, but rather must
be a meaningful summary of prior beliefs.  We note that difficulties
associated with prior specification evaporate when the models compared
are discrete with no obvious scientific models between; see
\citet[Section~6.5]{gelm:carl:ster:rubi:95} for examples and
discussion.  \cite{kass:raft:95} offers a thoughtful review of Bayes
factors including numerous examples, computational methods, and
methods for investigating sensitivity to the prior distribution.

Another popular method for model selection is the Bayesian Information
Criterion (BIC) \citep{schw:78}. BIC aims to select the model with the
highest posterior probability,
\begin{equation}
\hbox{choose the model that maximizes} \quad \pi_i \int
p(\bY|{\btheta}_i,{\rm Model}~i)p({\btheta}_1|{\rm
Model}~i)d{\btheta}_i,
\label{eq:maxpprob}
\end{equation}
where $\pi_i$ is the prior probability of model $i$. Since
computations of this posterior probability is generally complicated,
Schwartz suggested replacing it with BIC which maximizes the
loglikelihood with penalty term $-0.5 q_i\log n$, where $q_i$ is the
number of free parameters in model $i$ and $n$ is the sample
size. This penalty term favors models with fewer free parameters.
Schwartz went on to show that under certain conditions BIC is
asymptotically equivalent to using Equation~\ref{eq:maxpprob}. Like
the Bayes factor, the posterior probability in
Equation~\ref{eq:maxpprob} is highly sensitive to the choice of prior,
including the prior probabilities of the various models. That BIC is
not at all sensitive to the choice of prior distribution reflects the
fact that it is a poor approximation of Equation~\ref{eq:maxpprob}, at
least for small samples.
% and thus the use of BIC is devoid of its stated justification.

\section{An Fe~K Line in GRB 970508?}

In this section we examine the X-ray spectrum of the afterglow of
GRB~970508, analyzed for Fe~K line emission by \citet{piro:etal:99}.
This is a difficult and extremely important measurement: the detection
of X-ray afterglows from $\gamma$-ray bursts
%is at best a tricky business, relying
relies on near-real-time satellite response to unpredictable events
and a great deal of luck in catching a burst bright enough for a
useful spectral analysis.  The ultimate physics of these events is
still controversial, but they are among the most distant observable
objects in the sky.  Detecting a clear atomic (or cyclotron) line in
the generally smooth and featureless afterglow (or burst) emission not
only gives one of the few very specific keys to the physics local to
the emission region, but also provides clues or confirmation of its
distance (via redshift).

\citet{piro:etal:99} used the $F$ statistic to determine the
significance of the detected Fe~K line, as is standard practice.
Regularity conditions for the $F$ statistic and the LRT statistic are
similar; neither can be used to test whether the true parameter value
is on the boundary of the parameter space.  Thus, when testing for a
spectral line, the $F$ test is equally inappropriate. {\it We
emphasize that we do not highlight any errors particular to
\citet{piro:etal:99} but rather illustrate that the standard method
for identifying spectral lines is not trustworthy.} In fact, our more
rigorous analysis confirms their detection, but with higher
significance.

This section is divided into two parts. First we describe the
afterglow data, our models with and without an Fe~K emission line,
and two model fitting strategies, i.e., $\chi^2$ fitting in XSPEC and
Bayesian posterior analysis. Second we use posterior predictive
p-values to evaluate the strength of the evidence for the Fe~K
emission line.

\subsection{Data and Model Fitting}

An X-ray point-like emission associated with GRB~970508 was observed
by the BeppoSAX only six hours after the initial $\gamma$-ray burst
onset. The exposure time of 28ks was long enough to monitor the
evolving X-ray spectrum with an average observed flux of order $\sim$
$10^{-12}$~ergs~$\hbox{cm}^{-2}$~$\hbox{s}^{-1}$ in the combined
low-energy concentrator spectrometer (LECS) and medium-energy energy
concentrator spectrometer (MECS) instruments.  The data are plotted in
Figure~\ref{fig:raw:data}.  \citet{piro:etal:99} divided the data into
two time intervals ``1a'' versus ``1b'' and indicated that an emission
line, a possible redshifted Fe-K line, can only be present during the
initial phase of the observation,\footnote{\citet{piro:etal:99} were
specifically interested in this hypothesis which they proposed after
preliminary data analysis.  See Section 2 in \citet{piro:etal:99}.}
i.e., in the data set 1a.

We extracted the data from the BeppoSAX
archive\footnote{http://www.sdc.asi.it} in order to investigate the
significance of the line.  We restrict attention to data sets 1a LECS,
1a MECS, 1b LECS, and 1b MECS and use the default instrument responses
and background files provided for this observation. We fix the
relative normalization of LECS versus MECS at 0.8
\citep[c.f.,][]{piro:etal:99}, extract the spectra from the original
event files, and fit two models:
\begin{description}
\item[Model 1]: a simple absorbed power law;
\item[Model 2]: a simple absorbed power law and
               a Gaussian line at 3.5 keV with a
               known width, 0.5 keV (cf., \citet{piro:etal:99}).
\end{description}

Both models were fit via both $\chi^2$ fitting and Bayesian posterior
analysis. For $\chi^2$ fitting, we use XSPEC v.10 after binning the
data so that there are at least 15 counts (source+background) per bin;
the results are summarized in Table~\ref{tbl:xspec:fit}\footnote{The
best-fitted values we report for the power law fit (Model 1) are
within the intervals defined by error bars in Table 1 (i.e.,
Observation 1a+1b) in \citet{piro:etal:99}. That the two fits are not
identical is probably due to differences in binning mechanisms used,
to the fact that we fixed the relative normalization LECS/MECS at 0.8
and thus did not fit it, and to possible slight differences in
background files used. We also note that the best fitted line
intensity (see Model 2) we obtained agrees with that reported by
\citet{piro:etal:99} ($I_{\hbox{Fe}} =(5 +/- 2) \times 10^{-5}$
photons $\hbox{cm}^{-2}~\hbox{s}^{-1}$).} and yield an $F$ statistic
for comparing Model~1 and Model~2, i.e., $\Delta\chi^2/\chi^2_\nu$, of
4.156673.\footnote{\label{foot:diff:data} The data sets 1a and 1b LECS
and MECS are obtained by splitting two event files (for LECS and MECS)
that correspond to data sets 1 LECS and 1 MECS into an initial and a
later period. These split versions used by \citet{piro:etal:99} were
not available to us, so we did extractions from the event files
ourselves based on the information presented in Table 1 in
\citet{piro:etal:99}. The event files were retrieved from the BeppoSAX
archive at http://www.sdc.asi.it. We call our extractions 1a and 1b
LECS and MECS, but these files may not be exactly the same as the
split data sets analyzed by \citet{piro:etal:99}.  More specifically,
the exposure times for the four data sets we created are as follows:
1a LECS: 7.885 ks, 1b LECS: 9.361 ks, 1a MECS: 10.369 ks, 1b MECS:
17.884 ks.} Because the necessary regularity conditions are not met,
it is impossible to calibrate this $F$ statistic using the nominal $F$
reference distribution.

\if1\aastex\placefigure{fig:post:fit}\fi

To simulate from the posterior distribution of the parameters of
Model~1, we use the Markov chain Monte Carlo method of
\citet{vand:etal:01}; results appear in Figure~\ref{fig:post:fit}.
The Bayesian analysis requires us to specify prior distributions on
the model parameters. \citep[See][for details of Bayesian spectral
analysis.]{vand:etal:01} As illustrated in Figure~\ref{fig:like:post},
we have tried to use relatively non-informative prior
distributions.\footnote{We imposed a flat prior on the normalization
parameter, independent Gaussian relatively diffuse priors on photon
index and nH ($N(2, \sigma=.61)$ and $N(6, \sigma=2.1197)$
respectively), and for Model~2 a gamma prior on $\lamc$, $p(\lamc)
\propto \hbox{exp}(-0.05\lamc)$. } Figure~\ref{fig:like:post} compares
the marginal posterior distribution of the power law and photon
absorption parameters \footnote{The marginal posterior distribution of
the photon index, $\beta$, and photon absorption, $\gamma_{ph}$, is
obtained from Model~1 by integrating out the normalization parameter,
$\alpha$. I.e,.  $p(\beta, \gamma_{ph} |x, \hbox{no line is present})=
\int p({\btheta}| x, \hbox{no line is present}) \hbox{d}\alpha$.
Computing the marginal posterior distribution requires numerical
integration of the joint posterior distribution and is a
computationally expensive task. On a dual-processor 200Mhz Ultra-2 Sun
workstation with 256M RAM it took the program 183994.27 CPU time (in
seconds) to run.  We do not advocate such computation in general, but
rather use the marginal posterior distribution to verify the method in
this specific case.} with their marginal
likelihood\footnote{The marginal likelihood of the photon index,
$\beta$, and the photon absorption ,$\gamma_{ph}$, is obtained from
the likelihood function $p(x|\alpha, \beta, \gamma_{ph}, \hbox{no line
is present})$ by integrating out the normalization parameter,
$\alpha$, over a large, but finite interval containing the range of
physically meaningful values of $\alpha$.}  and illustrates the
similarity between the two, i.e., the non-informative nature of the
prior distribution.

\if1\aastex\placefigure{fig:like:post}\fi

The treatment of background in out Bayesian posterior analysis differs
from that of XSPEC. Instead of subtracting the background, we first
fit a power law to the background counts and then model the observed
counts as Poisson with intensity equal to the sum of the source and
fitted background intensities \citep[see][]{vand:etal:01}.  To
simulate from the posterior predictive distribution, we ran a Markov
chain Monte Carlo algorithm, discarding the first 100 simulations and
using every fifteenth simulation thereafter to obtain a total of 2000
simulations.  Convergence was judged by comparing the simulations to
contours of the posterior distribution; see
Figure~\ref{fig:like:post}. We do not advocate this as a general
strategy because of the computational effort required to obtain the
contours of the posterior distribution.  A more general method based
on multiple chains in described in \citet{gelm:rubi:92}; see also
\citet{vand:etal:01}.

\subsection{Calibrating the LRT Accounting for Parameter Uncertainty}

To evaluate the reference distribution of a test statistic, posterior
predictive analysis accounts for uncertainty in the parameter values
in the null model by sampling the parameters from their Bayesian
posterior distribution. Thus, following the four-step procedure
described in Section~4.1, we first simulate the parameters from their
posterior distribution as described in Section~5.1 (Step~1).  For each
simulation of the parameter, the 1a LECS and MECS and 1b LECS and MECS
are then simulated (Step~2) and the LRT statistic is computed
(Step~3).  The necessary maximum likelihood estimates are computed
with the EM algorithm as described by \citet{vand:00}.  The resulting
reference distribution is compared with the observed value of the LRT
statistic in Figure~\ref{fig:xspec:ppp}(a), yielding a posterior
predictive p-value of 0.75\% (Step~4).  Thus, the observed LRT
statistic is very unusual in the absence of the Fe~K line and we
conclude there is evidence for the Fe~K line, which is calibrated by
the magnitude of the posterior predictive p-values.

\if1\aastex\placefigure{fig:xspec:ppp}\fi

We can repeat this procedure using the posterior distribution under
Model~2 to simulate the parameters in Step~1.  That is, we can treat
the model with the emission line as the null model. The resulting
p-value (65.71\%) indicates no evidence against Model 2. In this
simulation, the data are generated {\it with} a spectral line that is
exactly the best fit line for the observed data. Thus, since the LRT
statistic is designed to detect this very spectral line as discussed
in Section~4.3, it is no surprise that the model is found to agree
with the data.  As we shall see, this situation persists; test
statistics that are not well designed to detect discrepancies between
the model and the data will never result in p-values that make us
question the null model.

The LRT statistic and $F$ statistic are not the only quantities that
can be used in posterior predictive checking. In fact, posterior
predictive p-values can be used to access how well a particular
important feature of the data is explained by a model. For example,
another measure of the strength the Fe~K line is the maximum
likelihood estimate of its intensity, $\lamc$.\footnote{We define
$\lamc$ as the rate per second of counts due to the Fe~K line for the
MECS instrument before absorption with the maximum effective area.}
Having already simulated from the posterior distribution of the model
parameters and the corresponding simulated data sets, we need only
compute the maximum likelihood estimate of $\tilde {\cal F}$ for each
data set to construct the reference distribution for $\lamc$.  The
results for both the model with and without the Fe~K line appear in
Figure~\ref{fig:xspec:ppp}(c)-(d).  Qualitatively, the conclusion
drawn from Figure \ref{fig:xspec:ppp}(c) is the same as with the LRT
statistic; the data is explained better by the model with the Fe~K
line.

If the continuum parameters (or more generally the parameters of the
null model) are {\it very well} constrained, we can perform an
approximate calibration of a test statistic by fixing the parameters
of the null model at their fitted values rather than simulating them
from their posterior distribution. That is, we can mimic the
simulations described in Section~3.2, using the ``parametric
bootstrap''.  We illustrate this procedure by calibrating the $F$
statistic calculated in Section~5.1. Specifically, we use a simulated
reference distribution that we obtain by simulating $N$ data sets
under a fully specified null model, (i.e., Model 1) with parameters
fixed at the best-fitted values in Table~\ref{tbl:xspec:fit}. The
simulations have the following steps:

\begin{enumerate}
\item
  Simulate $N$ data sets ({\tt fakeit} in XSPEC v.10)
  according to the null model with parameters fixed at their best-fit
  values, with the same effective area, instrument response, and
  background files as well as the exposure time as the initial
  analysis (i.e. simulate $\tilde{\bx}^{(t)} \dist p(\bx|{\btheta})$, for
  $t=1,\ldots,N$). Each simulated data set is appropriately binned, using
  the same binning algorithm as that used to bin the real data.
\item Fit the null and alternative model
  (i.e., Models~1 and 2) to each of the $N$ data sets and compute the
  $F$ statistic, $\Delta\chi^2/\chi^2_\nu$ for $t=1,\ldots,N$.
\item Compute the approximate p-value,
  $$p={1 \over N} \sum_{t=1}^N {\cal I}\{T_F(\tilde{\bx}^{(t)}) > T_F
  (\bx)\},$$
  where $T_F(\bx)$ represents the $F$ statistic.
\end{enumerate}

We emphasize that the simulation results rely on the assumption that
the data were actually generated under the null model with parameters
fixed at their best fit values. A probability histogram of the
simulated $F$ statistics can be used to calibrate the $F$ statistic
and compute a p-value; see Figure \ref{fig:xspec:ppp}(e).  The value
of the $F$ statistic is the 93.77th percentile of $F(1,13)$
distribution (i.e., $F$ distribution with degrees of freedom 1 and 13:
$F(1,13)=\chi_1^2/(\chi^2_{13}/13)$).  Thus, the nominal p-value is
6.23\%. The simulation, however, gives somewhat stronger evidence
against the null model reporting a p-value of 3.8\%.  Unfortunately,
this calibration is contingent on the accuracy of the model used to
simulate data in Step~1.

\section{Statistics: Handle with Care}
Although the LRT is a valuable statistical tool with many
astrophysical applications, it is not a universal solution for model
selection.  In particular, when testing a model on the boundary of the
parameter space (e.g., testing for a spectral line), the (asymptotic)
distribution of the LRT statistic is unknown.  Using this LRT and its
nominal $\chi^2$ distribution can lead to unpredictable results (e.g.,
false positive rates varying from 1.5\% to 31.5\% in the nominal 5\%
false positive rate test in Monte Carlo studies).  Thus, the LRT
should not be used for such model selection tasks.

The lesson to be learned from misapplication of the LRT is that there
is no replacement for an appreciation of the subtleties involved in
any statistical method.  Practitioners of statistical methods are
forever searching for statistical ``black boxes'': \ put the data in
and out pops a p-value or a fitted model.  When working with the
sophisticated models that are common in spectral, spatial, or temporal
analysis as well as other applications in astrophysics, such black
boxes simply do not exist.  The highly hierarchical structures
inherent in the data must be, at some level, reflected in the
statistical model.  Stochastic aspects of instrument response,
background counts, absorption, pile-up, and the relationship between
spectral, temporal, and spatial data must be accounted for.  With such
structured models, over-simplified, off the-shelf methods such as
assuming Gaussian errors (e.g., $\chi^2$ fitting) lead to
unpredictable results.  Standard
tests, such as the LRT, Cash, or $F$ statistics, sometimes are
appropriate (e.g., testing whether a mean parameter is equal to a
specified value) and sometimes are not (e.g., testing for the presence
of a spectral line) and are never appropriate with small data sets.

Even more sophisticated methods can have pitfalls.  Frequentist
methods (e.g., maximum likelihood with asymptotic frequentist error
bars), for example, typically rely on large sample sizes which may not
be justifiable in practice (e.g., for mixture models).  Non-parametric
methods require fewer parametric assumptions but often grossly
simplify the structure of the underlying model, discarding scientific
information.  Sometimes even these methods require strong assumptions
such as knowing the underlying model completely (e.g.,
Kolmogorov-Smirnov goodness-of fit tests).  Bayesian methods easily
accommodate the hierarchical structure in the data and do not require
asymptotic (large data set or many measurements) approximations.  The
computational tools required for highly structural models, however,
require careful implementation and monitoring; determining
convergence of Markov chain Monte Carlo methods is rarely automatic.
Moreover, although Bayesian statistical summaries (e.g., error bars)
are mathematically consistent summaries of information, they may not
exhibit the frequentist properties (e.g., coverage rates) that might
be expected.  Nevertheless, the computational difficulties that
sometimes exist with Bayesian analysis are much easier to overcome
than the conceptual difficulties that may arise in other frameworks
(e.g., unknown sampling distributions of test statistics).  Thus,
Bayesian methods are best equipped to handle highly structured models,
but we emphasize that like any statistical method, they must be used
with knowledge, sophistication, and care.

\if1\aastex \acknowledgements
    The authors gratefully acknowledge
    funding for this project partially provided by NSF grants
    DMS-97-05157 and DMS-01-04129, and
    by NASA Contract NAS8-39073 (CXC). \fi

\bigskip

\bigskip

\appendix
\noindent{\bf \huge{Appendices}}

\section{Regularity Conditions for the LRT}

Here we state the regularity conditions required for the standard
asymptotic behavior of the LRT. (Our presentation follows Chapter~4 of
\citet{serf:80} which should be consulted for details.)  Let $X_1,
... , X_n$ be independent identically distributed random variables
with distribution $F(x;{\btheta})$ belonging to a family ${\cal
F}=\{F(x;{\btheta}), {\btheta} \in \Theta \}$, where $\Theta \subset
{\cal R}^k$ is open and ${\btheta}=(\theta_1,...,
\theta_k)$. $F(x;{\btheta})$ are assumed to possess densities or mass
functions $f(x;{\btheta})$, that satisfy

\begin{enumerate}
\item For each ${\btheta} \in \Theta$, each $i=1,...,k$, each $j=1,...,k$,
and each $l=1,...,k$, the derivatives
  \begin{equation}
    \label{eq:R1}
g    \frac{\pd \hbox{log} f(x; {\btheta})}{\pd \theta_i}, \quad
    \frac{\pd^2 \hbox{log} f(x; {\btheta})}{\pd \theta_i \pd \theta_j}, \quad
    \frac{\pd^3 \hbox{log} f(x; {\btheta})}{\pd \theta_i \pd \theta_j \pd \theta_l}
  \end{equation}
  exist, all $x$;\footnote{Implicit here is the requirement that the
  support of the distribution be independent of $\theta$, otherwise
  there would be a $\theta$ and an $x$ for which the derivatives
  in (\ref{eq:R1}) would not exist.}
\item For each ${\btheta}_{*}\in \Theta$, there exist functions
$g(x)$, $h(x)$, and $H(x)$ (possibly depending on ${\btheta}_{*}$)
such that for ${\btheta}$ in a neighborhood $N({\btheta}_{*})\subset
\Theta$ the relations
    \begin{equation}
    \left| \frac{\pd  f(x; {\btheta})}{\pd \theta_i} \right| \le g(x), \quad
    \left|\frac{\pd^2 f(x; {\btheta})}{\pd \theta_i \pd \theta_j}\right| \le h(x), \quad
    \left|\frac{\pd^3 \hbox{log} f(x; {\btheta})}{\pd \theta_i \pd \theta_j \pd \theta_l} \right| \le
H(x)
  \end{equation}
  hold, all $x$ and all $1 \le i,j,l \le k$, with
  \begin{equation}
    \int g(x) dx < \infty, \int h(x) dx < \infty, \hbox{ and }
    \int H(x)f(x;{\btheta}) dx < \infty
    \hbox{ for } {\btheta} \in N({\btheta}_{*})
  \end{equation}
  \item For each ${\btheta} \in \Theta$ the information matrix
    \begin{equation}
       {\bf I}({\btheta})=\left[ E\left\{\frac{\pd \hbox{log} f(x; {\btheta})}{\pd \theta_i}
       \frac{\pd \hbox{log} f(x; {\btheta})}{\pd \theta_j} {\Big |} {\btheta} \right\} \right]_{k
\times k}
    \end{equation}
   exists and is positive definite.
\end{enumerate}
Consider $\Theta_0 \subset \Theta$ such that the specification of
$\Theta_0$ may be equivalently given as a transformation
\begin{equation}
   \theta_1=g_1(\nu_1, ..., \nu_{k-r}),
   ~...~ ,~
   \theta_k=g_k(\nu_1, ..., \nu_{k-r}),
\label{eq:reparam}
\end{equation}
where $\nu=(\nu_1, ... , \nu_{k-r})$ ranges through an open
  set $N\subset {\cal R}^{k-r}$. E.g., if $k=3$ and
  $\Theta_0=\{{\btheta}: \theta_1=\theta_1^{*}\}$, we then may take
  $N=\{(\nu_1, \nu_2): (\theta_1^{*}, \nu_1, \nu_2) \in \Theta_0 \}$
  and the functions $g_1, g_2, g_3$ to be
\[
   g_1(\nu_1,\nu_2)=\theta_1^{*}, ~
   g_2(\nu_1,\nu_2)=\nu_1, ~
   g_3(\nu_1,\nu_2)=\nu_2.
\]
Assume further that $g_i$ possess continuous first order partial
derivatives and that
\[
    {\bf D_\nu }= \left[ \frac{\pd g_i}{\pd \nu_j} \right]_{k \times (k-r)}
\]
is of rank $k-r$. Alternatively, if $\Theta_0$ is defined by a set of
$r$ ($r\le k$) restrictions given by equations \[ R_i({\btheta})=0,
\quad 1\le i \le r \] (e.g., in the case of a {\it simple} hypothesis
$\Theta_0=\{(\theta_1^{0}, \theta_2^{0}, \theta_3^{0})\}$, we have
$R_i({\btheta})=\theta_i-\theta_i^{0}$, $i=1,2,3$), we require that
$R_i({\btheta})$ possess continuous first order derivatives and that
\[
   {\bf C_{\btheta}}=\left[\frac{\pd R_i}{\pd \theta_j} \right]_{r\times k}
\]
is of rank $r$. Let ${\btheta}^{*}\in\Theta$ denote the true unknown
value of the parameter ${\btheta}$.  Define the null hypothesis to be
$H_0: {\btheta}^{*} \in \Theta_0$. Then if $H_0$ is true, the LRT
statistic (see Equation \ref{eq:defLRTa}) is asymptotically
distributed as $\chi^2$ with $r$ degrees of freedom.

\section{Testing for Lines: A Misapplication of the LRT}
\label{subseq:testing:lines}
Under the required regularity conditions, the asymptotic $\chi^2$
distribution of the LRT is based on the asymptotic normality of the
maximum likelihood estimate\footnote{The maximum likelihood estimate
  is a statistic with a sampling distribution. A theorem in
  mathematical statistics establishes that under the same regularity
  conditions required for the LRT to be asymptotically $\chi^2$ the
  distribution of the maximum likelihood estimate becomes Gaussian
  (i.e., Normal) as the sample size increases; see \citet{serf:80}.}
with mean equal to the true parameter value. If the true value is on
the boundary of the parameter space, however, the mean of the maximum
likelihood estimate cannot possibly be the true value of the
parameter since maximum likelihood estimates are always in the
parameter space.  (Clearly $\E(\hat {\btheta} \vert {\btheta} )$ can not be
${\btheta}$ if $\hat {\btheta}$ is, for example, always greater than
${\btheta}$.)  Thus, one of the regularity conditions required for the
LRT is that $\Theta$, the parameter space, be an open set.  This is
not the case when we test for the presence of a component in a finite
mixture model.  Consider the source model given in
Equation~\ref{eq:deflam}.  If we set ${\btheta}=(\alpha, \beta, \gamma,
\tilde{\cal F})$ and test $\tilde{\cal F}=0$ we are examining the
boundary of the parameter space, and the LRT statistic is not (even
asymptotically) $\chi^2$ in distribution.

We illustrate this point again using the model specified by Equation
\ref{eq:deflam}.  For simplicity we assume in this example that there
is no background or instrument response, effective area and absorption
are constant, and $(\alpha, \beta, \gamma)$ are fixed. Thus, the only
unknown parameter in Equation~\ref{eq:deflam} is the intensity
$\tilde{{\cal F}}$.  Our goal is to test whether the data are
consistent with the null model, i.e., $\tilde{{\cal F}}\in
\Theta_0=\{\tilde{{\cal F}}=0\}$ or if there is evidence for the more
general alternative model, i.e., $\tilde{{\cal F}}\in
\Theta=\{\tilde{{\cal F}}\ge 0\}$.  Let $Y_j$ denote the counts in
channel $j$ and let $\tau$ be the exposure time, then $ P(Y_j \vert
{\cal F}_j \tau) = (\exp(-{\cal F} \tau)) ({\cal F}_j \tau)^{Y_{j}} /
(Y_{j}!) $, or $Y_j\sim \hbox{Poisson}({\cal F}_j \tau)$ and the
log-likelihood is given by
\begin{equation}
\label{eq:loglike}
l(\tilde{{\cal F}}|\bY)=\sum_{j=1}^J \left[Y_j\hbox{log}({\cal F}_j \tau)
-{\cal F}_j \tau\right]
\end{equation}
where ${\cal F}_j$ is given in Equation~\ref{eq:deflam}.
The first derivative with respect to $\tilde{{\cal F}}$ is given by
\begin{equation}
\label{eq:loglikeprime}
l'(\tilde{{\cal F}}|\bY)=\sum_{j=1}^J
\left[Y_j\frac{p_j}{{\cal F}_j}
-p_j \tau\right],
\end{equation}
where $\bY=(Y_1,...,Y_J)$ and $p_j$ is as in Equation~\ref{eq:deflam}.
Since
\begin{equation}
l''(\tilde{{\cal F}}|\bY)=-\sum_{j=1}^J
Y_j\frac{p_j^2}{{\cal F}_j^2},
\end{equation}
$l''(\tilde{{\cal F}}|\bY)\le0$, and
$l'(\tilde{{\cal F}}|\bY)$ is a decreasing function of
$\tilde{{\cal F}}$.  If
\begin{equation}
 l'(\tilde{{\cal F}}=0|Y)\le0,
\label{eq:lprimecond}
\end{equation}
(i.e. the data happen to fluctuate so that the first derivative $\le
0$) the maximum likelihood estimate for $\tilde{{\cal F}}$ must be $0$
(see Figure~\ref{fig:lambda:example}).  In this case the LRT
statistic, $-2\hbox{log}R(\bY)$ must be $0$ since the maximum
likelihood estimate under the null and the alternative models are both
$0$ and thus $R(\bY)=1$. To compute the false positive rate of the LRT
we assume there is no spectral line (i.e., $\lamc=0$ as in the null
model) and compute the probability that we reject the null model in
favor of the alternative model.  According to the central limit
theorem, the distribution of $Y_j$ converges to Gaussian as $\tau$
increases. Since the expectation of the first term of the summand in
Equation \ref{eq:loglikeprime} is equal to the second term,
$l'(\tilde{{\cal F}}|\bY)$ converges to a Gaussian with mean $0$ as
$\tau$ increases. Therefore asymptotically the LRT statistic equals
zero $50\%$ of the time, whence the reference distribution of the LRT
statistic cannot be a $\chi^2$ distribution with any number of degrees
of freedom.  Examples of this kind are well known to statisticians and
in this case it can be shown that the distribution of the LRT
statistic is itself a {\it mixture} distribution taking on the value
zero with probability 50\% and follows a $\chi^2_1$ with probability
50\% (e.g., see Section 5.4 in \cite{titt:smit:mako:85}).
\cite{mattox:etal:96} notice this same behavior in a Monte Carlo
evaluation of the null distribution of the LRT statistics for testing
for point source in Energetic Gamma Ray Experiment Telescope data.
This mixture distribution should be used to calibrate the LRT
statistic when testing for a single line when all other parameters are
fixed.

\if1\aastex\placefigure{fig:lambda:example}\fi

In practice, this may mean that in cases where the continuum is
extremely well constrained by the data, and the width and position of
the possible line are known, the LRT or $F$ test could underestimate
the true significance by about a factor of two.  But there is no
guarantee that this will occur in real data; particularly when the
continuum is not well constrained the true significance can be
underestimated or overestimated.

\if1\btex\bibliography{../../../../TeXfiles/BibTeX/my}\else

\fi

\if1\preprint

\begin{center}
\begin{figure}
    \includegraphics[width=6.5in,height=3in]{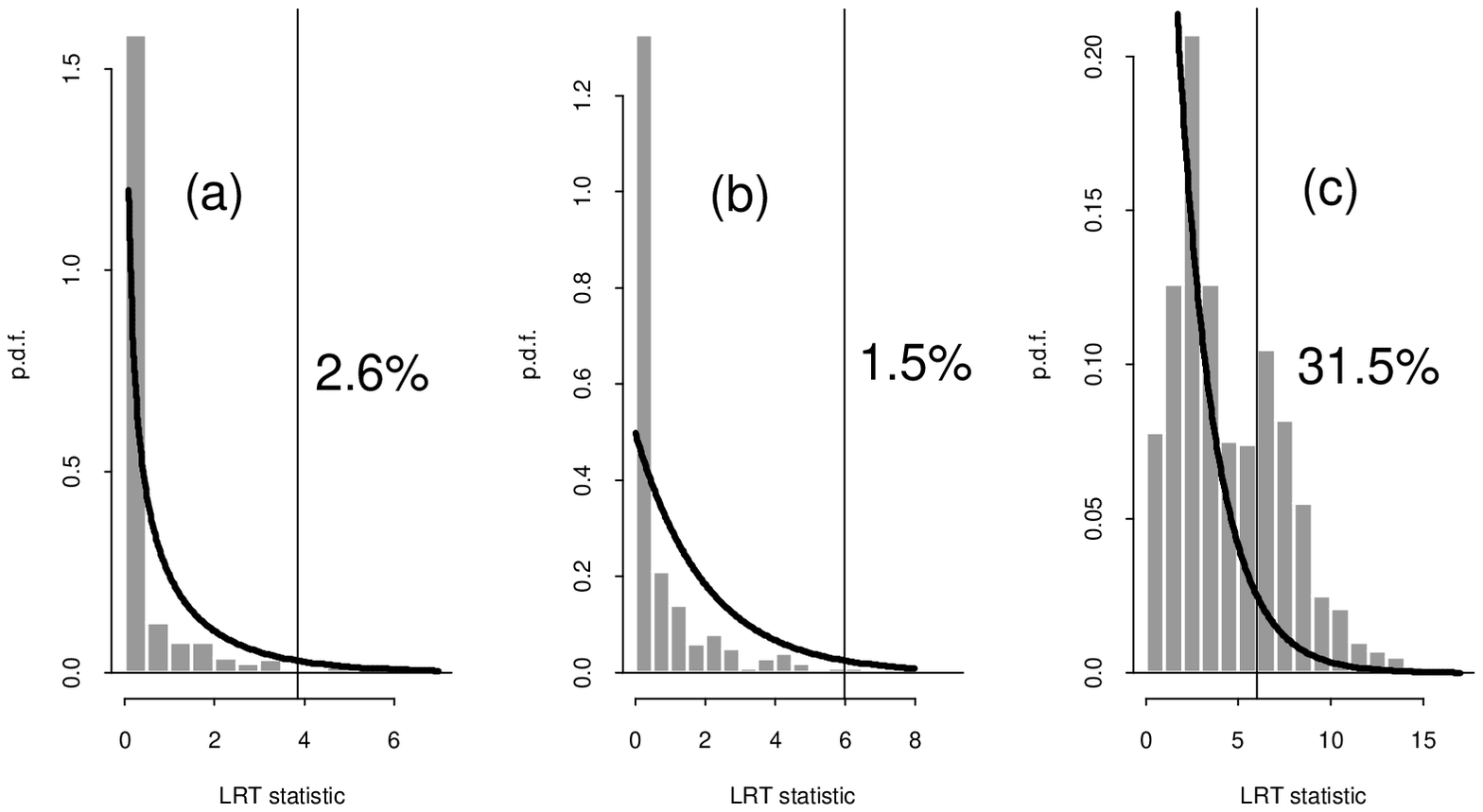}
    \caption{The null distribution of the LRT test statistic. The
      histograms illustrate the simulated null distribution of the LRT
      statistic in three scenarios and should be compared with nominal
      $\chi^2$ distributions which are also plotted. As detailed in
      Section~3.2, histogram (a) corresponds to testing for a narrow
      emission line with fixed location, histogram (b) corresponds to
      testing for a wide emission line with fitted location, and
      histogram (c) corresponds to testing for an absorption line.
      The vertical lines show the nominal cut off for a test with a
      5\% false positive rate; note the actual false positive rates
      vary greatly at 2.6\%, 1.5\%, and 31.5\%. The label on the
      y-axis stands for the probability density function (p.d.f.).
  \label{fig:lrt}}
\end{figure}
\end{center}

\begin{center}
\begin{figure}
    \includegraphics[width=6 in,height=3in]{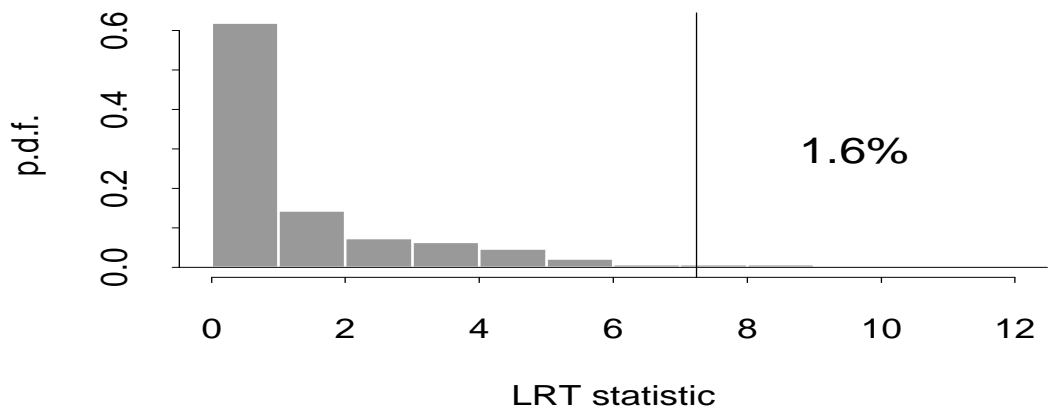}
    \caption{The posterior predictive distribution of the LRT test
      statistic. This histogram gives the expected variation under
      $p(T(\tilde x)\cur|x)$ for the quasar image. Note the observed
      value, $T(x)=7.24$, giving evidence against the null (no
      spectral line) model.
  \label{fig:ppp}}
\end{figure}
\end{center}

\begin{center}
\begin{figure}
   \includegraphics[width=5.6in, height=6in]{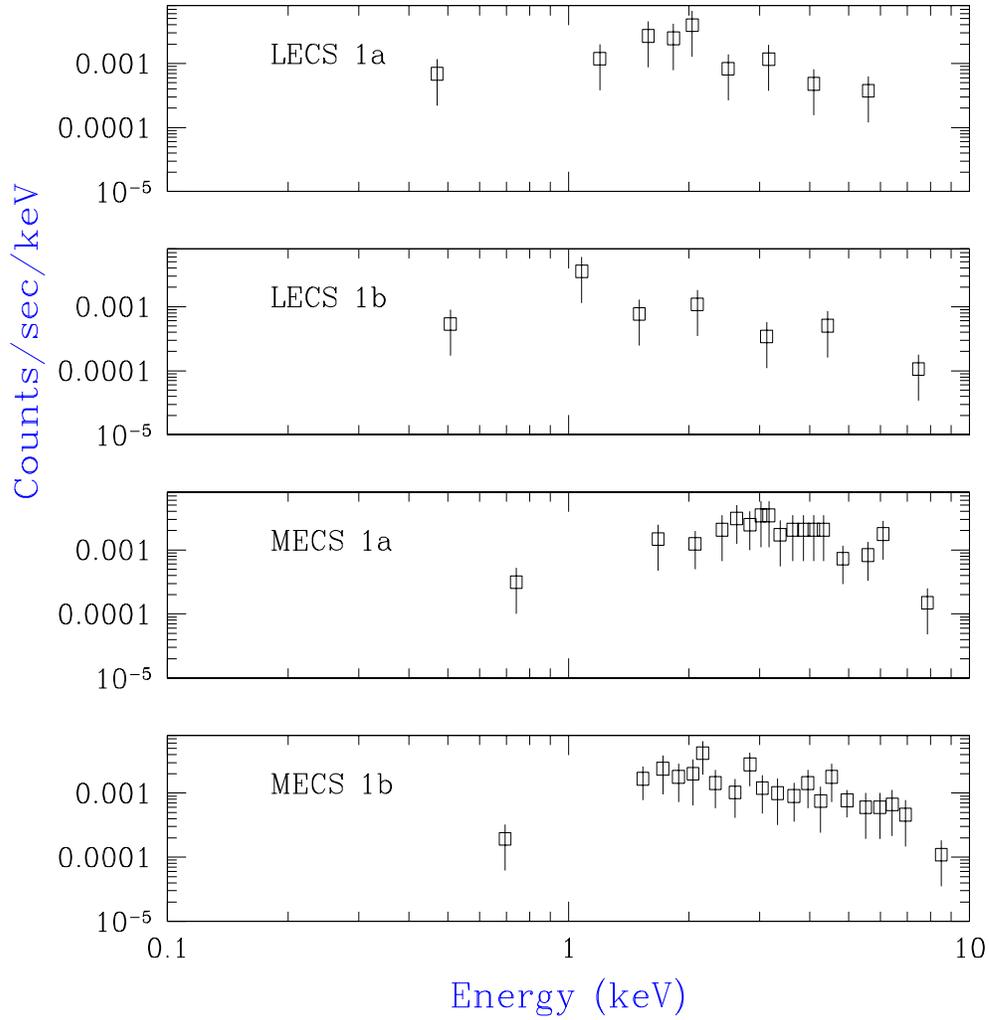}
    \caption{The GRB 970508 observation.  The four panels
      corresponding to the four data sets: LECS 1a \& 1b and MECS 1a
      \& 1b. The posited spectral line is at 3.5 keV with width of
      .5keV in LECS 1a and MECS 1a; it is not present in LECS 1b and
      MECS 1b.  
\label{fig:raw:data}}
\end{figure}
\end{center}

\begin{center}
\begin{figure}
   \includegraphics[width=5.6in, height=6in]{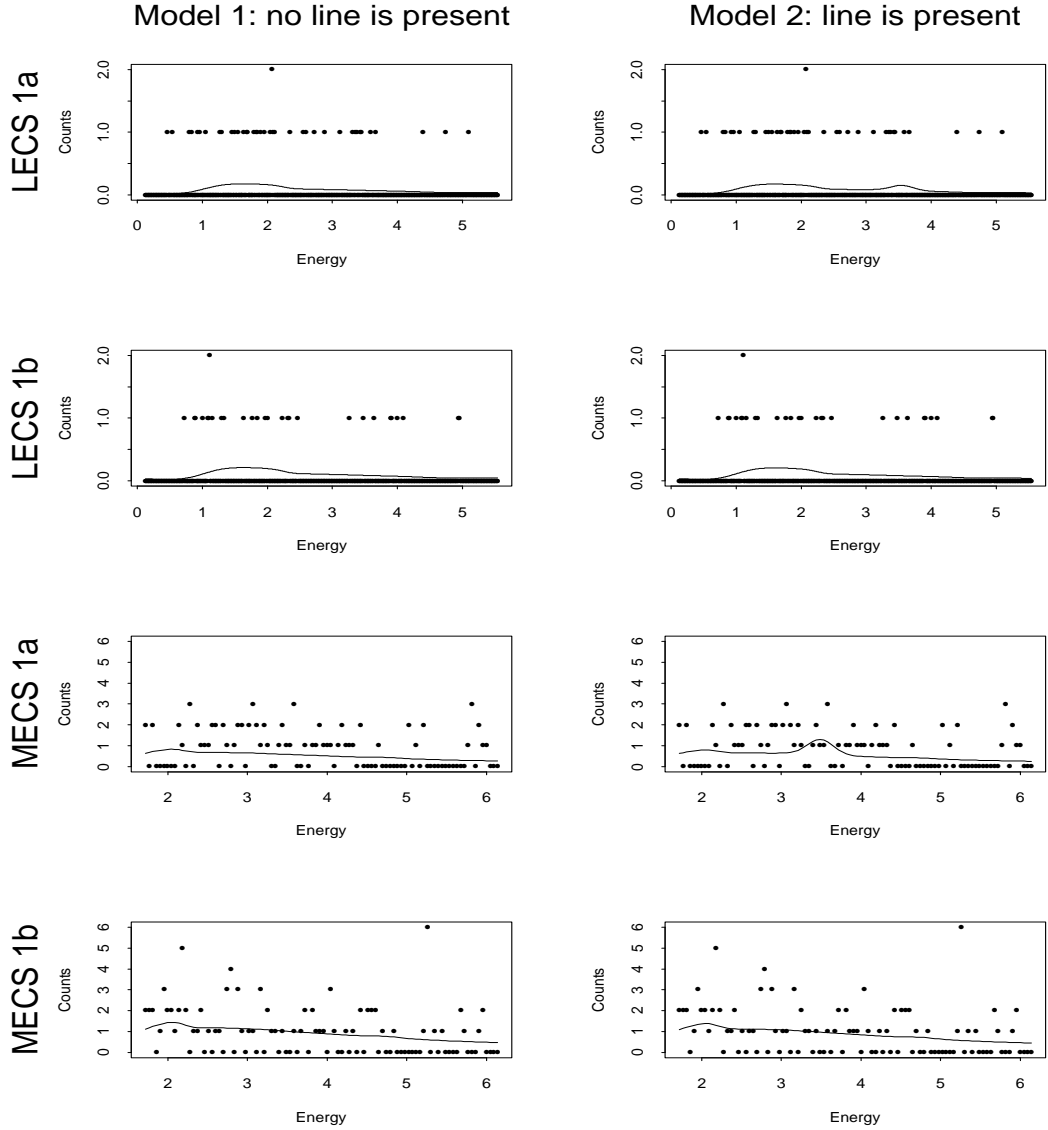}
    \caption{Posterior Analysis of GRB 970508. The black dots
    correspond to the counts registered in the pha channels. There are
    four data sets: LECS 1a \& 1b and MECS 1a \& 1b.  Model 1 stands
    for a simple power law with photon absorption and Model 2 for a
    simple power law with photon absorption and a Gaussian line at 3.5
    keV with fixed width .5keV.  Model 2 assumes further that the line
    is only present in LECS 1a and MECS 1a, but not in LECS 1b or MECS
    1b.  The posterior distribution of the parameters for each model
    was studied separately using the method of van Dyk
    \emph{et~al.}(2001). The curves in the plots represent the fitted
    expected counts and were computed by fixing parameters in each
    model at their posterior means.  
\label{fig:post:fit}}
\end{figure}
\end{center}

\begin{center}
\begin{figure}
    \includegraphics[width=5.6in,height=2.6in]{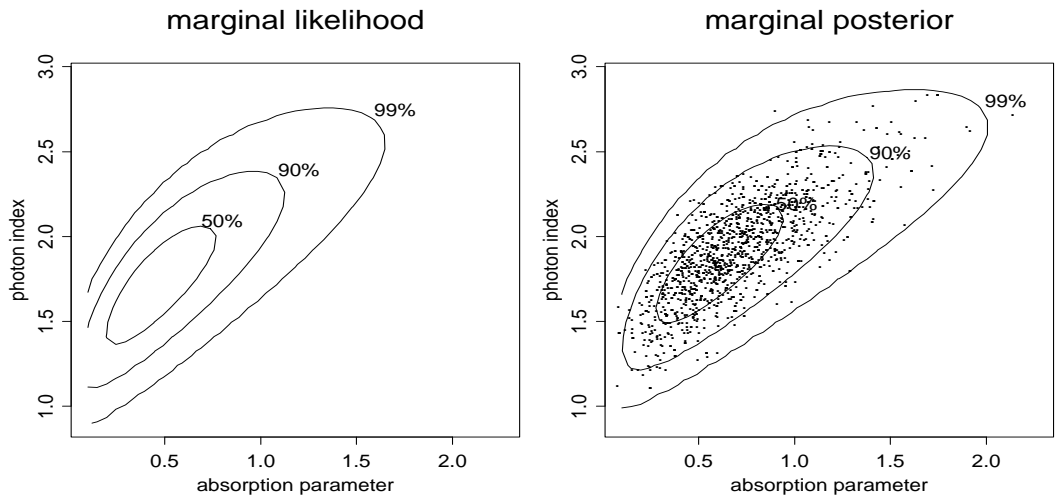}
    \caption{Contours of the marginal likelihood and marginal
     posterior distribution. The contours corresponds to 50\%, 90\%,
     and 99\% of the area under each surface; in the case of the
     marginal posterior distribution, these corresponds to posterior
     probabilities. Comparing the two plots illustrates that the prior
     distribution is relatively uninformative. In the plot of the
     marginal posterior distribution we compare the numerically
     computed contours with 1500 Monte Carlo simulations generated
     with the Gibbs sampler; the Monte Carlo simulations are displayed
     as points and follow the contours well.
\label{fig:like:post}}
\end{figure}
\end{center}

\begin{center}
\begin{figure}
   \includegraphics[width=5.6in]{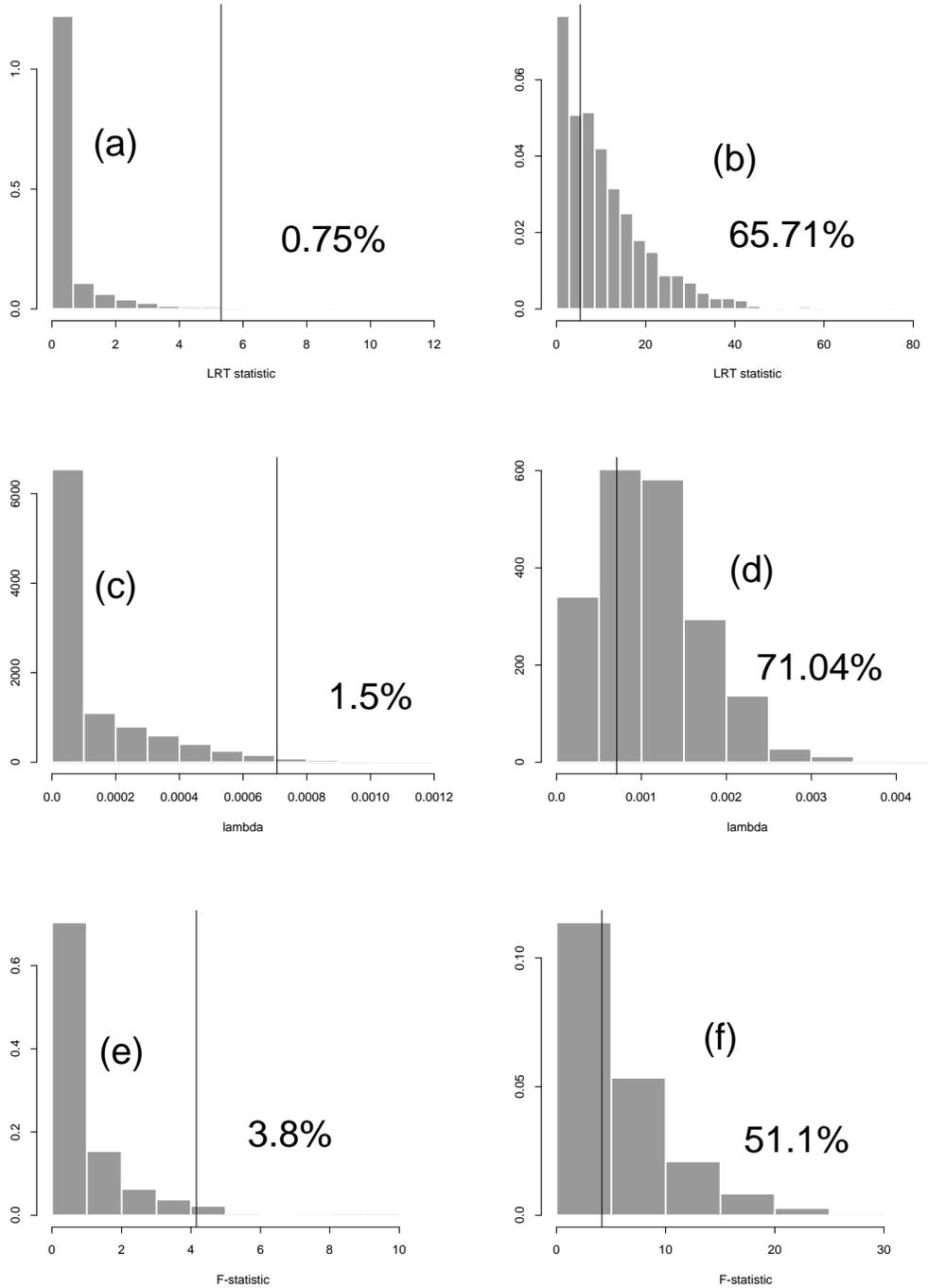} 
   \caption{ Analysis
   of GRB 970508.  {\bf (a)-(b)}: 2000 simulations from (a) $p(T_{\rm
   LRT}(\tilde x)|x,\hbox{no line is present})$ and (b) $p(T_{\rm
   LRT}(\tilde x)|x, \hbox{line is present})$ were used to produce
   these probability histograms of the posterior predictive
   distributions for the LRT statistic.  Percentages indicate the mass
   of the corresponding distribution to the right of the vertical line
   at 5.3, the observed value of the LRT statistic.  {\bf (c)-(d):}
   Observed value of the maximum likelihood estimate for $\lamc$ at
   0.0007 against probability histograms of 2000 simulations from the
   posterior predictive distribution of this quantity under 2 models:
   (c) without a spectral line and (d) with a spectral line.  We give
   percentages of the mass of the distribution to the right of
   the vertical line at 0.0007. Histogram (c) illustrates that the
   model with a spectral line is clearly preferable.  {\bf (e)-(f):}
   Observed value of the $F$ statistic (vertical line at 4.2) against
   probability histograms of 1000 simulations obtained from null
   distributions under two fully specified models: (e) model 1 with
   parameters as they appear in Table~\ref{tbl:xspec:fit} and (f)
   model 2 with parameters as they appear in
   Table~\ref{tbl:xspec:fit}. Percentages indicate the mass of the
   corresponding distribution to the right of 4.2.
\label{fig:xspec:ppp}}
\end{figure}
\end{center}

\begin{center}
\begin{figure}
    \includegraphics[width=6.5in,height=3in]{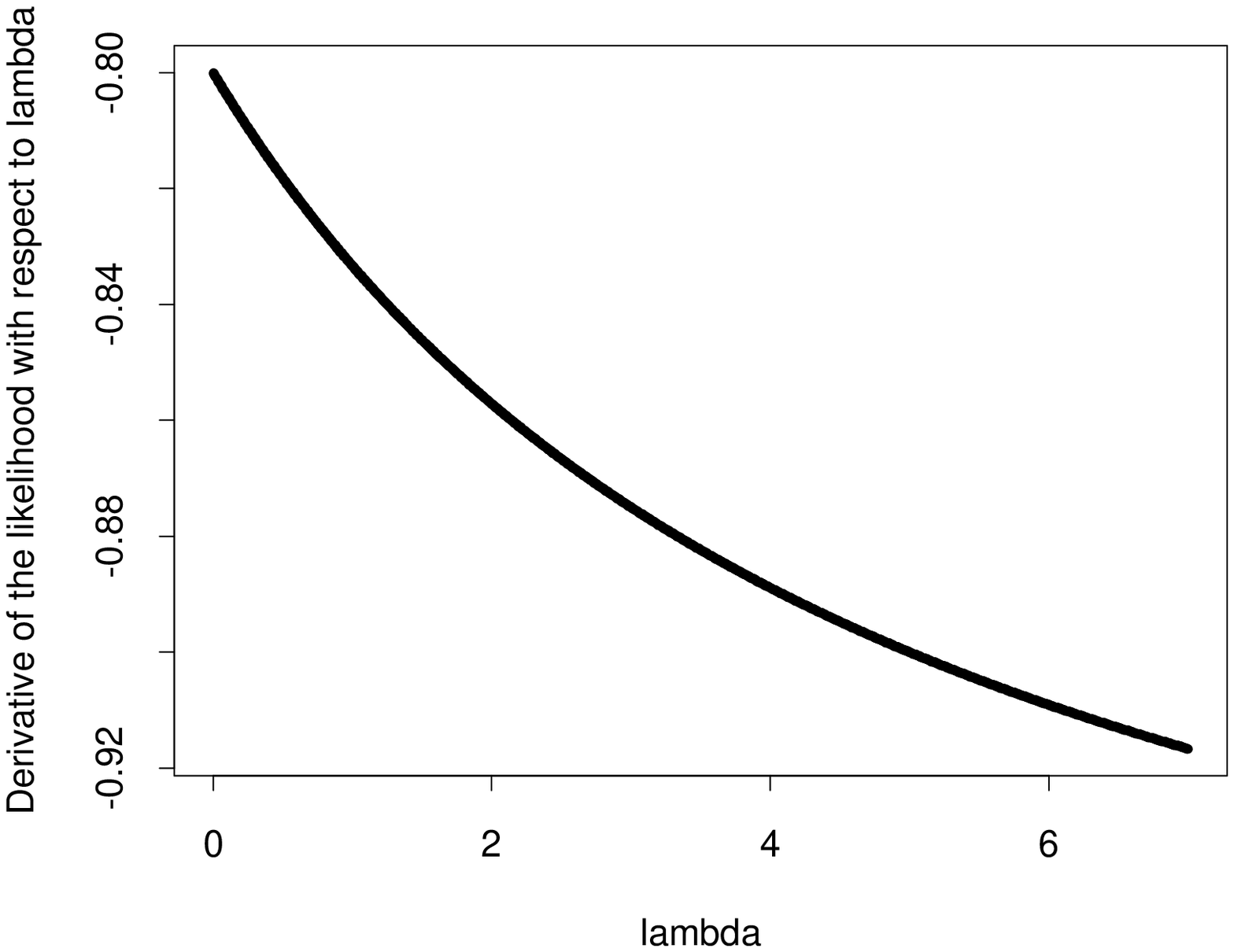}
    \caption{ If $l'(\tilde{{\cal F}}=0|Y)\le0$, then $l'(\tilde{{\cal
    F}}|Y)\le0$ for any $\tilde{{\cal F}} \ge 0$ since
    $l'(\tilde{{\cal F}}|Y)$ is a monotonically decreasing function of
    $\tilde{{\cal F}}$, recall: $l''(\tilde{{\cal F}}|Y) \le0$ for all
    $\tilde{{\cal F}}$ and $Y$. But this implies that $l(\tilde{{\cal
    F}}|Y)$ is maximized at $\tilde{{\cal F}}=0$ whenever
    $l'(\tilde{{\cal F}}=0|Y)\le0$.  This figure is a qualitative
    illustration.  
\label{fig:lambda:example}}
\end{figure}
\end{center}

\fi

\if0\preprint

\begin{figure}
    \caption{The null distribution of the LRT test statistic. The
      histograms illustrate the simulated null distribution of the LRT
      statistic in three scenarios and should be compared with nominal
      $\chi^2$ distributions which are also plotted. As detailed in
      Section~3.2, histogram (a) corresponds to testing for a narrow
      emission line with fixed location, histogram (b) corresponds to
      testing for a wide emission line with fitted location, and
      histogram (c) corresponds to testing for an absorption line.
      The vertical lines show the nominal cut off for a test with a
      5\% false positive rate; note the actual false positive rates
      vary greatly at 2.6\%, 1.5\%, and 31.5\%. The label on the
      y-axis stands for the probability density function (p.d.f.).
  \label{fig:lrt}}
\end{figure}

\begin{figure}
    \caption{The posterior predictive distribution of the LRT test
      statistic. This histogram gives the expected variation under
      $p(T(\tilde x)\cur|x)$ for the quasar image. Note the observed
      value, $T(x)=7.24$, giving evidence against the null (no
      spectral line) model.
  \label{fig:ppp}}
\end{figure}

\begin{figure}
    \caption{The GRB 970508 observation.  The four panels
      corresponding to the four data sets: LECS 1a \& 1b and MECS 1a
      \& 1b. The posited spectral line is at 3.5 keV with width of
      .5keV in LECS 1a and MECS 1a; it is not present in LECS 1b and
      MECS 1b.  
\label{fig:raw:data}}
\end{figure}

\begin{figure}
    \caption{Posterior Analysis of GRB 970508. The black dots
    correspond to the counts registered in the pha channels. There are
    four data sets: LECS 1a \& 1b and MECS 1a \& 1b.  Model 1 stands
    for a simple power law with photon absorption and Model 2 for a
    simple power law with photon absorption and a Gaussian line at 3.5
    keV with fixed width .5keV.  Model 2 assumes further that the line
    is only present in LECS 1a and MECS 1a, but not in LECS 1b or MECS
    1b.  The posterior distribution of the parameters for each model
    was studied separately using the method of van Dyk
    \emph{et~al.}(2001). The curves in the plots represent the fitted
    expected counts and were computed by fixing parameters in each
    model at their posterior means.  
\label{fig:post:fit}}
\end{figure}

\begin{figure}
    \caption{Contours of the marginal likelihood and marginal
     posterior distribution. The contours corresponds to 50\%, 90\%,
     and 99\% of the area under each surface; in the case of the
     marginal posterior distribution, these corresponds to posterior
     probabilities. Comparing the two plots illustrates that the prior
     distribution is relatively uninformative. In the plot of the
     marginal posterior distribution we compare the numerically
     computed contours with 1500 Monte Carlo simulations generated
     with the Gibbs sampler; the Monte Carlo simulations are displayed
     as points and follow the contours well.
\label{fig:like:post}}
\end{figure}

\begin{figure}
   \caption{ Analysis
   of GRB 970508.  {\bf (a)-(b)}: 2000 simulations from (a) $p(T_{\rm
   LRT}(\tilde x)|x,\hbox{no line is present})$ and (b) $p(T_{\rm
   LRT}(\tilde x)|x, \hbox{line is present})$ were used to produce
   these probability histograms of the posterior predictive
   distributions for the LRT statistic.  Percentages indicate the mass
   of the corresponding distribution to the right of the vertical line
   at 5.3, the observed value of the LRT statistic.  {\bf (c)-(d):}
   Observed value of the maximum likelihood estimate for $\lamc$ at
   0.0007 against probability histograms of 2000 simulations from the
   posterior predictive distribution of this quantity under 2 models:
   (c) without a spectral line and (d) with a spectral line.  We give
   percentages of the mass of the distribution to the right of
   the vertical line at 0.0007. Histogram (c) illustrates that the
   model with a spectral line is clearly preferable.  {\bf (e)-(f):}
   Observed value of the $F$ statistic (vertical line at 4.2) against
   probability histograms of 1000 simulations obtained from null
   distributions under two fully specified models: (e) model 1 with
   parameters as they appear in Table~\ref{tbl:xspec:fit} and (f)
   model 2 with parameters as they appear in
   Table~\ref{tbl:xspec:fit}. Percentages indicate the mass of the
   corresponding distribution to the right of 4.2.
\label{fig:xspec:ppp}}
\end{figure}

\begin{figure}
    \includegraphics[width=6.5in,height=3in]{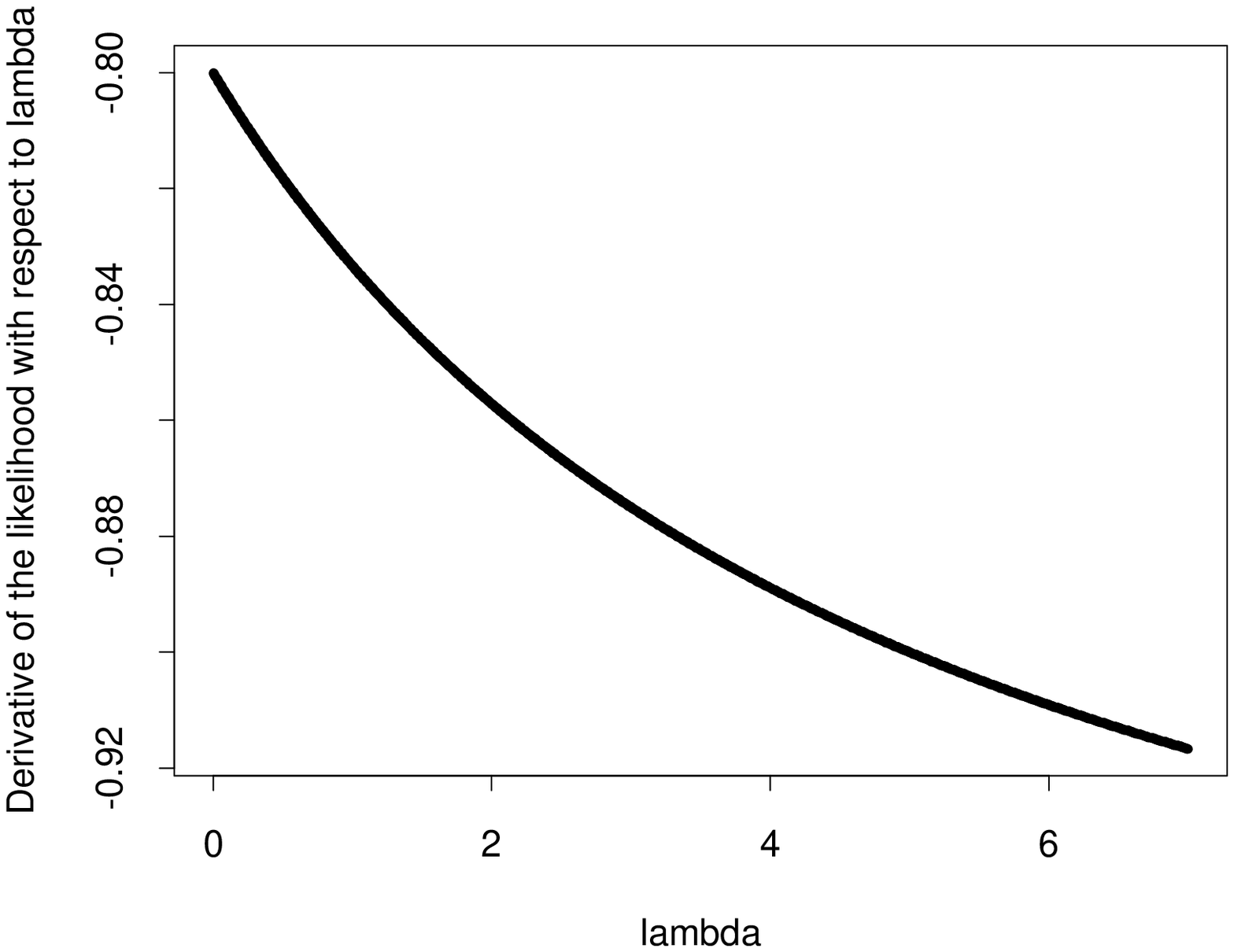}
    \caption{ If $l'(\tilde{{\cal F}}=0|Y)\le0$, then $l'(\tilde{{\cal
    F}}|Y)\le0$ for any $\tilde{{\cal F}} \ge 0$ since
    $l'(\tilde{{\cal F}}|Y)$ is a monotonically decreasing function of
    $\tilde{{\cal F}}$, recall: $l''(\tilde{{\cal F}}|Y) \le0$ for all
    $\tilde{{\cal F}}$ and $Y$. But this implies that $l(\tilde{{\cal
    F}}|Y)$ is maximized at $\tilde{{\cal F}}=0$ whenever
    $l'(\tilde{{\cal F}}=0|Y)\le0$.  This figure is a qualitative
    illustration.  
\label{fig:lambda:example}}
\end{figure}
\fi

%\begin{table}
%\begin{center}
%\begin{tabular}{lc}
%\hline\hline
%Type of Test &Number of Papers\\
%\hline
%Test for Spectral Emission Line & 43\\
%Test for Spectral Absorption Feature& 20\\
%Test for Other {\it Added} Spectral Feature & 26\\
%Test for Gaussian Feature in Light Curves& 1\\
%Test for Quasi-Periodic Oscillation in Timing Data& 11\\
%Test for Image Component & 5\\
%Comparing Non-Nested Models & 17\\
%Other Questionable Cases& 4\\
%Seemingly Appropriate Use of Test & 56\\
%\hline\hline
%\end{tabular}
%\caption{Results of a Survey of papers in ApJ, ApJL, and ApJS published
%between 1995 and mid 2001 and returned by a search of for `F
%statistic', `F test', or `LRT' at the ApJ website. Many
%papers contain multiple uses of the tests (especially among the first three
%rows in the table). Such papers only appear once.
%\label{tbl:search}}
%\end{center}
%\end{table}

\begin{table}[r]
\begin{center}
\begin{tabular}{ccccc}
\hline\hline
Model & norm & photon index & n$_{H}$ ($10^{22}$) & Gaussian norm \\\hline
{\bf Model 1}: No emission line & 3.6424E-04 & 1.928 & 0.6397 & {\bf n/a} \\
$\chi^2=25.74467$,  $\nu=14$ & & & & \\\hline
{\bf Model 2}: Emission line is &   4.3427E-04 & 2.186 &  0.70393 &  4.7633E-05 \\
present at 3.5keV (width 0.5 keV) & & & &  \\
$\chi^2=19.50732$,  $\nu=13$ & & & & \\
\hline\hline
\end{tabular}
\caption{ Minimal $\chi^2$ fits from XSPEC for data sets 1a+1b.
\label{tbl:xspec:fit}}
\end{center}
\end{table}

\end{document}